\documentclass[%
 reprint,
superscriptaddress,
 amsmath,amssymb,
 aps,
showkeys,
]{revtex4-2}

\usepackage{graphicx,xcolor,soul}
\usepackage{siunitx}
\usepackage[version=3]{mhchem}
\usepackage{bm}
\usepackage{multirow}
\usepackage{tabularx}

\usepackage{hyperref}
\usepackage{natbib}

\begin{document}

\preprint{MSSP}

\title{Environmental variation compensated damage classification and localization in ultrasonic guided wave SHM using self-learnt features and Gaussian mixture models}

\author{Shruti Sawant}
\affiliation{Department of Electrical Engineering, Indian Institute of Technology Bombay, Mumbai 400076, MH India}

\author{Sheetal Patil}
\affiliation{Department of Electrical Engineering, Indian Institute of Technology Bombay, Mumbai 400076, MH India}

\author{Jeslin Thalapil}
\affiliation{Department of Electrical Engineering, Indian Institute of Technology Bombay, Mumbai 400076, MH India}

\author{Sauvik Banerjee}
\email{sauvik@civil.iitb.ac.in}
\affiliation{Department of Civil Engineering, Indian Institute of Technology Bombay, Mumbai 400076, MH India}

\author{Siddharth Tallur}
\email{stallur@ee.iitb.ac.in}
\affiliation{Department of Electrical Engineering, Indian Institute of Technology Bombay, Mumbai 400076, MH India}


\date{\today}

\begin{abstract}
Conventional damage localization algorithms  used in ultrasonic guided wave-based structural health monitoring (GW-SHM) rely on physics-defined features of GW signals.  In addition to requiring domain knowledge of the interaction of various GW modes with various types of damages, they also suffer from errors due to variations in environmental and operating conditions (EOCs) in practical use cases. While several machine learning tools have been reported for EOC compensation, they need to be custom-designed for each combination of damage and structure due to their dependence on physics-defined feature extraction. Research on the use of deep learning tools such as convolutional neural networks (CNNs) for automated feature extraction in GW-SHM is in a nascent stage and has vast untapped potential for increasing the robustness of GW-SHM systems. In this work, we propose a CNN-based automated feature extraction framework coupled with Gaussian mixture model (GMM) based EOC compensation and damage classification and localization method. Features learnt by the CNNs are used for damage classification and localization of damage by modeling the probability distribution of the features using GMMs. The Kullback-Leibler (KL) divergence of these GMMs with respect to corresponding baseline GMMs are used as signal difference coefficients (SDCs) to compute damage indices (DIs) along various GW sensor paths, and thus for damage localization. The efficacy of the proposed method is demonstrated using FE generated GW-data for an aluminum plate with a network of six lead zirconate titanate (PZT) sensors, for three different types of damages (rivet hole, added mass, notch) at various temperatures (from \SI{0}{\degreeCelsius} to \SI{100}{\degreeCelsius}), with added white noise and pink noise to incorporate errors due to EOCs. We also present experimental validation of the method through characterization of notch damage in an aluminum panel under varying and non-uniform temperature profiles, using a portable custom-designed field programmable gate array (FPGA) based signal transduction and data acquisition system. We demonstrate that the method outperforms conventional EOC compensation method using GMM with physics-defined features for damage localization in GW-SHM systems prone to EOC variations. 
\end{abstract}

\keywords
{Convolutional neural networks (CNN), structural health monitoring (SHM), ultrasonic non-destructive testing (NDT), Gaussian mixture model (GMM), damage localization}
                              
\maketitle

\section{Introduction}
Structural health monitoring (SHM) systems employed for detecting defects in large infrastructure such as bridges, aircraft, railway tracks etc. consist of large number of sensors mounted on the structure to perform a variety of measurements to satisfy the vastly different requirements across different structures and applications. Ultrasonic guided wave-based SHM (GW-SHM) is popular approach for a variety of structures due to low cost of instrumentation, sensitivity of high frequency waves to small defects, ability of guided waves (GWs) to travel long distances on the structure without attenuation etc.\cite{mitra2016guided,yan2010ultrasonic,abbas2018structural} However, a significant practical limitation of damage detection algorithms in GW-SHM systems designed in laboratory environment, is susceptibility to errors introduced through variations in environmental and operational conditions (EOCs) such as temperature, moisture absorption, loading etc.\cite{attarian2014long,cawley2018structural,konstantinidis2006temperature,clarke2009evaluation,schubert2012influence,douglass2020model}
GW-based damage characterization requires interpreting damage features of time-series corresponding to GW signals recorded by sensor. Variation in EOCs, including temperature, vibration, loading, moisture etc. result in change in phase and amplitude of GW signals \cite{roy2015load,sikdar2018effects}. Identifying effective damage features is crucial to the success of damage characterization. Selection of suitable features usually requires professional knowledge and varies for different structures and types of damages. To relieve the burden of damage feature selection, researchers have been investigating deep learning algorithms to automate damage identification without explicit feature definition in advance.
There are number of studies reporting temperature compensation techniques \cite{croxford2010efficient,sohn2007effects}. These can be broadly classified into two classes: baseline signal stretch (BSS) and optimal baseline selection (OBS). In BSS, a single baseline signal (e.g. recorded under nominal operating conditions at room temperature) is stretched to match the signal obtained at different temperatures. In OBS approach, baseline signals are recorded at different temperatures and subsequent computations utilize selection of a baseline signal within the database which is most similar to the monitoring signal recorded during inspection. Other preprocessing techniques to remove the effect of EOC variation recently reported in literature include homoscedastic nonlinear cointegration \cite{zolna2016towards}, normalized smoothed envelope threshold (NSET) coupled with high pass filtering \cite{seno2019passive}, variational mode decomposition (VMD) algorithm for denoising and removing seasonal patterns in guided wave signals \cite{mousavi2021prediction} and cointegration analysis \cite{dao2014lamb,tome2020damage}.

Machine learning (ML) tools prove useful in modeling complex nonlinear phenomena particularly in the presence of uncertainty. In OBS approach, data-driven techniques are used, wherein training of a model is performed using data collected from prior experiments. In the absence of physical models, these non-parametric methods are useful for successful damage characterization. Data-driven techniques useful for such problems include principal component analysis (PCA) \cite{liu2015robust}, singular value decomposition (SVD) \cite{clarke2010guided} and independent component analysis (ICA) \cite{dobson2015independent}. Among OBS techniques, data driven probabilistic approaches are found to be most effective \cite{gorgin2020environmental}. 
Abdeljaber et al. used convolutional neural network (CNN) for vibration-based structural damage detection and localization \cite{ABDELJABER2017154}. Khan et al. demonstrated CNN classifier for analyzing delaminations in composite structures wherein CNNs are trained with 2-dimensional spectrograms obtained from time domain vibration signal using Short Time Fourier Transform \cite{KHAN2019586}. Bao et al. trained deep neural network using gray scale images obtained using time-series vibration signals \cite{Bao2019}. CNNs were trained using stacked autoencoders and greedy layer-wise training. To demonstrate the applicability of this architecture, data from a long-span cable-stayed bridge in China was used. In another work, Aria et al. applied deep learning to predict damage size and remaining life of the structure using acoustic emission measurements \cite{Aria2020}. Giannakeas et al. demonstrated a method to capture impact of EOC variation on Lamb waves in composites through the Bayesian calibration of a finite element (FE) model using experimental observations and compute the model assisted probability of detection (MAPOD) \cite{giannakeas2022digital}.

While deep learning based algorithms for automated feature extraction have been reported in vibration based SHM system, there are limited reports for GW-SHM.
Xu et al. extracted features from GW signals and these features were processed by a CNN designed and trained to further extract high-level features and implement feature fusion for crack evaluation \cite{s19163567}. Pandey et al. have demonstrated a local interpretable model-agnostic explanations (LIME) tool for 1D CNN model to study damages (simulated as well as experimental) in aluminum plates \cite{PANDEY2022108220}. 
%
Rautela et al. recently presented extensive performance evaluation of several deep learning approaches assisted with reduced-order spectral finite element models for damage characterization \cite{rautela2020ultrasonic}. 
Ren et al. proposed Gaussian Mixture Model (GMM) based model for compensation of time varying temperature \cite{ren2019gaussian,ren2019multi,ren2020gaussian}. This involves direct extraction of features using time-domain signal recordings, without performing additional transforms.
Another example of OBS method for non-uniform temperature gradient on experimental data acquired using sensor array placed on aluminum plate was presented by Sun et al. \cite{sun2019identification}.  
Wang et al. explored the effect of such coupling of two EOC parameters, namely loading and temperature, on GW characteristics \cite{wang2020improved}. They utilized compensation strategy based on matching pursuit algorithm wherein features calculated using amplitude and group velocity are used for training the model.
Khurjekar et al. have reported deep learning models for feature extraction and damage localization in GW-SHM system wherein the variations in environmental factors are modeled as uncertainties in the data \cite{khurjekar2020uncertainty}, albeit without any estimate of damage intensity. 
Liu et al. used convolutional neural network based framework to represent damage characterization problem as image classification problem \cite{liu2019deep}. There have been limited reports of using deep learning approaches for damage characterization, albeit in absence of variation in EOCs \cite{melville2018structural,liu2019deep}. It emerges that research on application of deep neural networks for EOC compensation has been limited and has extensive scope for performance evaluation and optimization.

\begin{figure*}[!t]
\centering
\includegraphics[width=0.7\linewidth]{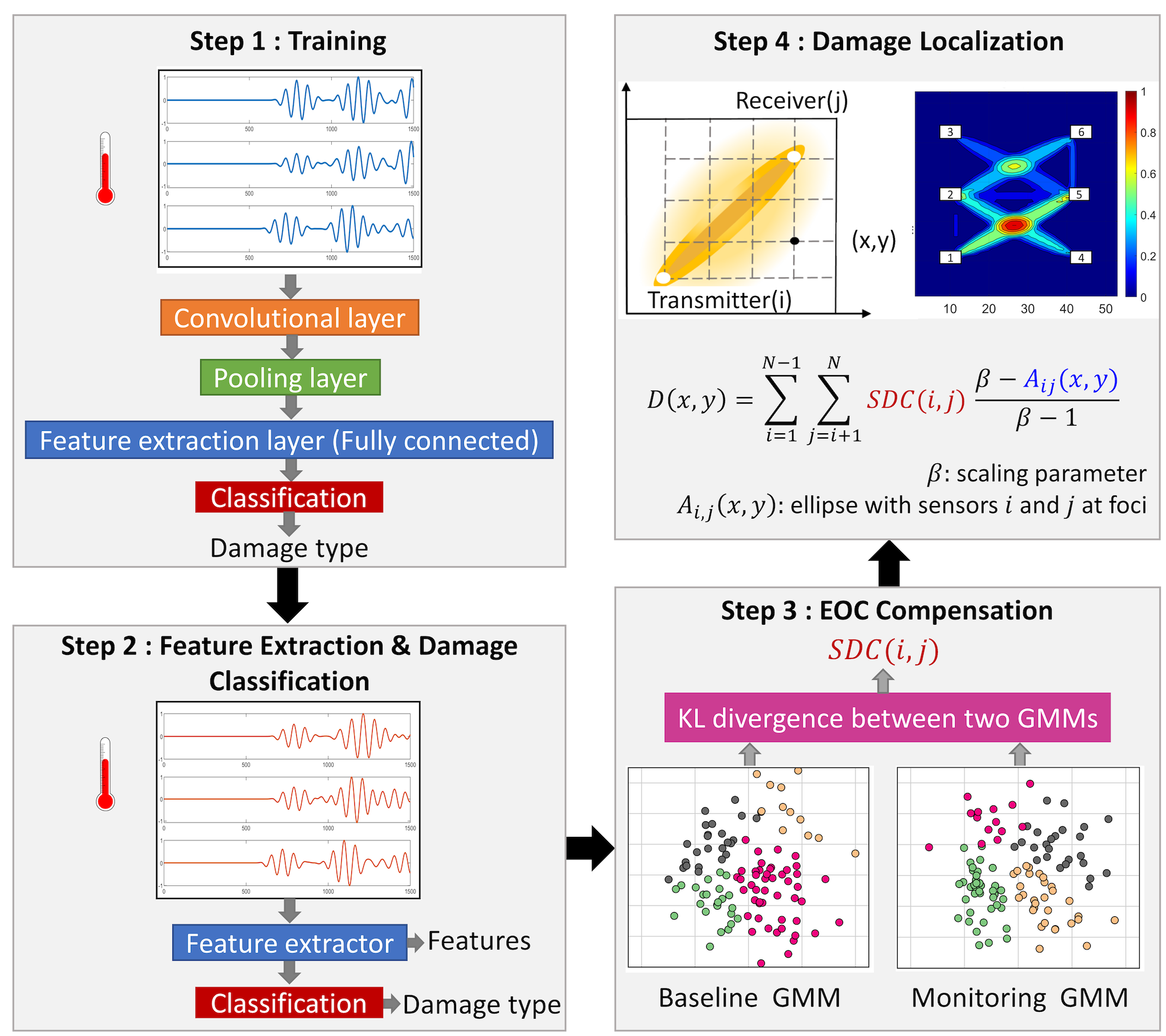}
\caption{Illustration of the proposed method using 1D-CNNs for feature extraction and GMMs for temperature compensated damage localization in GW-SHM systems.}
\label{fig:algo}
\end{figure*}

Conventional algorithms for damage localization use manually computed features based on time and/or frequency domain representation of the signals, and physics-based analysis of impact of damage and EOCs on signal properties (e.g. amplitude, phase, frequency etc.) of GW modes.
In this work, we propose a novel method for temperature compensated damage classification and localization using ultrasonic GWs.
The method utilizes CNNs to learn the features of GW signals acquired on healthy and damaged structures, and to classify the type of damage. The self-learnt features are then analyzed using GMMs for EOC compensation. The GMMs obtained using monitoring data are compared against those obtained with baseline data and the Kullback-Leibler (KL) divergence values of the GMMs are used as signal difference coefficient (SDC) for damage index assessment and damage localization.
The utility of this method is validated using GW data from a network of six lead zirconate titanate (PZT) sensors on an aluminum plate at various temperatures (from \SI{0}{\degreeCelsius} to \SI{100}{\degreeCelsius}) generated using FE simulations, with added white and pink noise of various signal to noise ratios (SNR: \SI{2}{dB}, \SI{4.5}{dB} and \SI{10}{dB}) and three types of damage (rivet hole, added mass and notch). Typically GW-SHM systems are designed for SNR of the order of \SI{20}{}-\SI{50}{dB} \cite{pedram2018split}. To maintain such high SNR, large values of actuation voltage are used for the transducers. We demonstrate that the proposed method works efficiently even for much lower values of SNR. 
Finally, we present experimental validation of this technique for notch damage in an aluminum panel using a custom-designed field programmable gate array (FPGA) based signal transduction and data acquisition system, and demonstrate that the method presented in this work outperforms conventional damage localization algorithms in presence of non-uniform and varying temperature conditions. The stochastic variation in the performance of the algorithm due to added noise is significantly reduced by employing the proposed method as compared to earlier reported approach for temperature compensation using manually computed features and GMM \cite{ren2019gaussian,ren2019multi}. For damage localization, features are extracted in automated fashion, without specifying any statistical relation or domain knowledge such as group velocity, choice of wave mode etc. The method presented here thus holds great promise approach for damage localization and characterization in GW-SHM systems prone to EOC variations.

\section{CNN+GMM based temperature compensated damage localization method}

The proposed method mainly consists of four steps as illustrated in Figure \ref{fig:algo}:

\begin{itemize}
  \item Training: Instead of manually computing features from time-domain signals, we employ a 1-dimensional (1D) CNN for damage classification and feature extraction using time-domain GW signals. GW signals are recorded for various sensor paths, denoted as pair of transmitter $i$ and receiver $j$. A CNN is set up for each path, and trained using data for baseline (healthy) and damaged conditions.
  
  \item Feature extraction \& damage classification: Once trained, the penultimate layer of the trained CNN is used for automated feature extraction and damage classification. This feature extraction layer is used to extract features from monitoring data obtained from the structure under maintenance. These features are used for damage classification by assigning labels to training data, as well as damage localization as described in the following steps. Note that several recordings are used for each sensor path in the monitoring stage, to emulate data obtained under various EOCs over various times.
  
  \item EOC compensation: Next, the features computed in the previous step are used to form clusters using GMMs. This step is crucial for compensating the effect of temperature and other uncertainties. The GMM for each path formed in monitoring stage is compared against the corresponding GMM set up with baseline data. The deviation of the monitoring GMM from baseline GMM is quantified using the KL divergence (i.e. relative entropy of the two probability distributions). 
  
  \item Damage localization: KL divergence of each path $i-j$ is used as the corresponding SDC to compute the damage index and perform damage localization using conventional GW SHM damage assessment method \cite{sikdar2016identification}. For a sensor network comprising of $N$ sensors, we can obtain $N \times (N-1)$ SDC values considering all possible sensor pairs, however not all paths may be necessary. 
\end{itemize}

\begin{figure}[!t]
\centering
\includegraphics[width=0.8\linewidth]{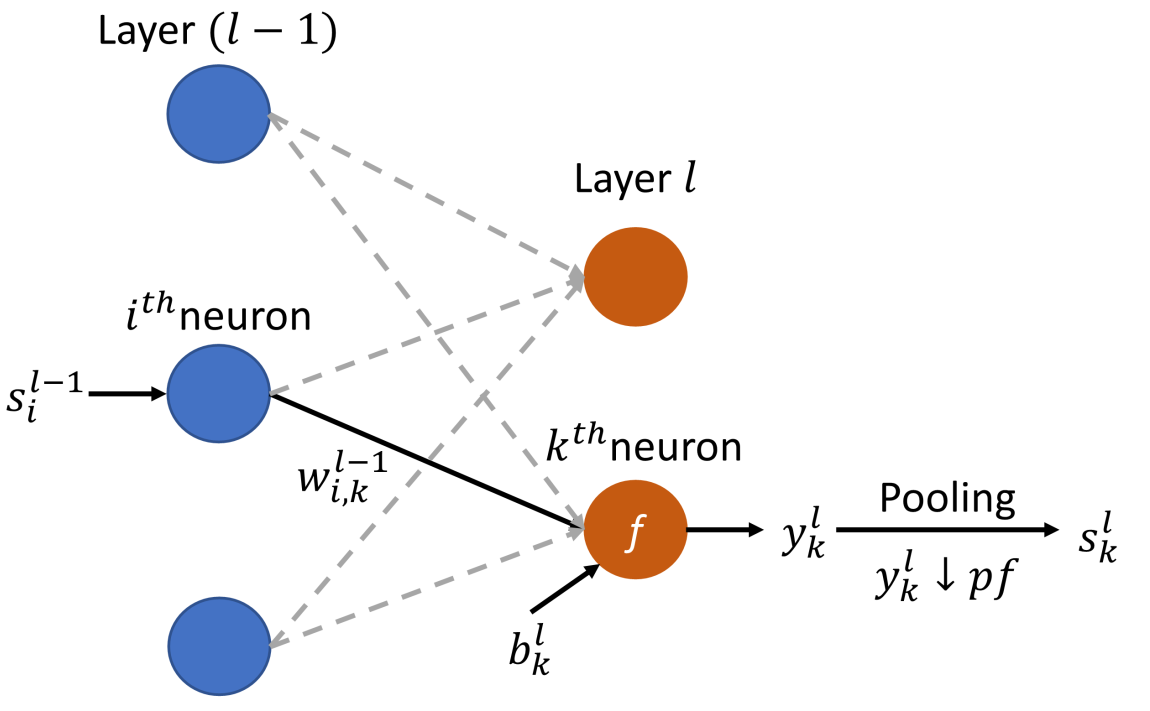}
\caption{Illustration of CNN architecture.}
\label{fig:CNN_GMM_illustration}
\end{figure}

\subsection{CNN based feature extraction}
Typical CNN architecture consists of convolutional layers, activation function, pooling layers and a fully connected layer. Convolution operation for 1D time-series data between neuron $i$ of layer $l-1$ and neuron $k$ of layer $l$ is defined as follows (Figure \ref{fig:CNN_GMM_illustration}): 

\begin{equation}
x_{k}^{l} = b_{k}^{l} + \sum conv(w_{ik}^{l-1}, s_{i}^{l-1})
\label{eq:conv}
\end{equation}

Here, $conv$ represents the convolution operation, $w_{ik}$ is the weight corresponding to the connection between neuron $i$ of layer $l-1$ and neuron $k$ of layer $l$, and $b_{k}^{l}$ is the bias of layer $l$. The result of this convolution operation $x_{k}^{l}$ is then processed by a nonlinear activation function $f$. Typical choices of nonlinear activation functions include rectified linear activation function (ReLU), sigmoid, tanh etc. The intermediate output $y_{k}^{l}$ of this activation function is expressed as:

\begin{equation}
y_{k}^{l} = f(x_{k}^{l})
\end{equation}

Next, pooling is performed to reduce dimensionality in order to reduce the number of trainable parameters and computational cost in the network, thereby controlling over-fitting. Popularly used pooling method is \textit{maxpooling} wherein, the highest-valued feature in a specified span is retained while other features in the specified span are discarded. Other pooling methods include average pooling or $L2-$norm pooling. 
All neurons in the fully connected layer $l$ have connections to activation function outputs from the preceding layer $l-1$.
The set of features of layer $l$ obtained through these operations is defined as:

\begin{equation}
s_{k}^{l} = y_{k}^{l} \downarrow pf
\label{eq:features}
\end{equation}

Here, $pf$ represents down-sampling factor used in the pooling operation. The feature set of the final layer of the CNN is used for classification, using the nonlinear sigmoid function to compute the probabilities of the input belonging to various classes (e.g. healthy, or various types of damage such as rivet hole, added mass, notch etc.) Equations \eqref{eq:conv}-\eqref{eq:features} describe the \textit{forward pass} in CNN training. During back propagation step of training CNNs, a cost function is computed using estimated targets and true targets. The network is then iteratively optimised using stochastic gradient descent to obtain optimal values of weights and biases for given time-series dataset (inputs). The training is deemed complete when the cost function saturates, or reduces below a pre-determined threshold. Once training is completed, these optimised weights and biases are used directly for feature extraction, without having to manually compute features using physics-based methods. The penultimate (fully connected) layer of the trained CNN computes features as defined in equation \eqref{eq:features}, and is therefore used as the feature extractor layer shown in Figure \ref{fig:algo}. 
Suppose an $n-$dimensional time-series is given as input to the CNN, we obtain an $m-$dimensional feature vector at the output of the feature extractor layer ($m << n$). These features are used as inputs for the next step of the method i.e. clustering using GMM for EOC compensation. 

\subsection{GMM based EOC compensation}

GMM is soft probabilistic clustering model that describes the membership of sample to a set of clusters using a mixture of Gaussian densities \cite{reynolds2009gaussian}. The probabilistic nature of GMM makes it a suitable choice for compensating uncertainties introduced by variations in EOCs. Unlike hard clustering, this method assigns probabilities to a sample belonging to a finite number of Gaussian distributions with unknown parameters. The optimum number of Gaussian components $k$ is determined using \textit{silhouette criterion}. The silhouette score of the GMM takes into account following two components:
\begin{itemize}
    \item Mean distance between a sample and all other samples in the same cluster
    \item Mean distance between a sample and all other samples in neighbouring cluster
\end{itemize}

The silhouette score varies in the range $[-1, 1]$.  A larger value of silhouette score indicates well defined clusters. 
Suppose for baseline data (GMM with $k$ clusters), the Gaussian components are represented as:

\begin{equation}
p(x) = \sum_{i=1}^{i=k} \pi_i \mathcal{N}(x|\mu_i,{\sigma_i})
\end{equation}

Here, the model parameters $\mu_i$, $\sigma_i$ and $\pi_i$ denote the mean, standard deviation and weight of $i^{th}$ Gaussian component. These parameters are obtained using expectation maximisation (EM) algorithm. The probability distribution function (PDF) $p(x)$ for feature $x$ captures the probabilistic nature of baseline time-series data obtained under various EOCs, and represented by features obtained from the CNN. Similarly, the PDF of monitoring data $q(x)$ is also obtained. After obtaining PDFs for both baseline and monitoring data, we need to quantify their mutual difference using an appropriate statistical relation. This is calculated using KL divergence \cite{goldberger2003efficient}, defined as:

\begin{equation}
    KL(p||q) = - \sum_x x*p(x)*\log\left[ \frac{q(x)}{p(x)}\right]
\label{eq:KL}
\end{equation}

Intuitively, if the probability of an event $x$ from distribution $p$ is significantly different from that of one from distribution $q$, then the resulting KL divergence magnitude will be large. In case of SHM application, if the monitoring data deviates substantially from baseline data (due to damage), the resulting KL divergence for that sensor path will be large. Therefore, the KL divergence value is used as SDC for computing damage index (DI) and performing damage localization.  

\subsection{Damage localization algorithm}
This subsection presents a brief overview of the SDC based DI map computation originally presented by Sikdar et al. \cite{sikdar2016identification}. 
For the arrangement shown in step 4 of Figure~\ref{fig:algo}, $(x_i,y_i)$ and $(x_j,y_j)$ denote Cartesian coordinates of transmitter $(i)$ and receiver $(j)$, respectively. The distance between the transmitter and receiver is denoted as $d_{(i,j)}$. The aluminum panel is represented by a $2-$dimensional grid, and at each grid point $(x,y)$, the pixel value for DI map is computed by scaling the ellipse $A_{i,j}(x,y)$ with focii as the transmitter $(i)$ and receiver $(j)$, and corresponding SDC coefficient $SDC(i,j)$:

\begin{equation}
DI(x,y) = \sum_{i=1}^{N-1} \sum_{j=i+1}^{N} SDC(i,j) \left[\frac{\beta - A_{i,j}(x,y)}{\beta - 1}\right]
\label{eq:di}
\end{equation}
where $\beta$ is an empirically determined scaling parameter independent of velocity of the propagating wave. The elliptical contour shaped spatial distribution function with non-negative values, $A_{i,j}(x,y)$ is defined as:

\begin{equation}
A_{i,j}(x,y)=
\begin{cases}
  P_{(i,j)}(x,y), & P_{(i,j)}(x,y)<\beta 
\\
  \beta, & P_{(i,j)}(x,y)\geq \beta
\end{cases}
\end{equation}
where
\begin{multline}
    P_{(i,j)}(x,y) = \\
    \frac{\sqrt{(x-x_i)^2+(y-y_i)^2} + \sqrt{(x-x_j)^2+(y-y_j)^2}}{d_{(i,j)}}
\end{multline}

The composite DI map is obtained by overlaying ellipses $A_{i,j}(x,y)$ for various transmitter-receiver $(i-j)$ paths. Intersections of paths with damage present in-line generate pixels with high intensity in the DI map, and therefore are useful for damage localization. Higher contrast between values of SDCs for damaged paths and undamaged paths yields better localization results in the DI map.

\section{Generation of temperature-affected guided wave  dataset for simulation study}\label{sec:FEM_COMSOL}

Propagation of Lamb waves in a plate depends on the thickness, density and elastic material properties of the structure.
For an isotropic plate of thickness $h$, the Lamb wave dispersion equations are specified as follows\cite{Rose2014}:
\begin{itemize}
    \item Symmetric (S) mode:
        \begin{equation}
        \frac{\tanh{\frac{\beta h}{2}}}{\tanh{\frac{\alpha h}{2}}}=\frac{4k^2\alpha\beta}{\left(k^2+\beta^2\right)^2} 
        \label{eq:sym}
        \end{equation}
    \item Anti-symmetric (A) mode:
        \begin{equation}
        \frac{\tanh{\frac{\beta h}{2}}}{\tanh{\frac{\alpha h}{2}}}=\frac{\left(k^2+\beta^2\right)^2}{4k^2\alpha\beta} 
        \label{eq:asym}
        \end{equation}
\end{itemize}
where $\alpha = \sqrt{k^2-\frac{\omega^2}{V^2_P}}$ and $\beta = \sqrt{k^2-\frac{\omega^2}{V^2_S}}$, $k=\frac{\omega}{c_{p}}$ is the wave number, $\omega$ is the angular frequency of the propagating wave with a phase velocity $c_{p}$. $V_P$ and $V_S$ are the velocities of the pressure (P)-wave and shear (S)-wave respectively, expressed as \cite{Rose2014}:

\begin{equation}
    V_P=\sqrt{\frac{\lambda+2\mu}{\rho}}
\end{equation}
\begin{equation}
    V_S=\sqrt{\frac{\mu}{\rho}}
\end{equation}
The terms $\lambda$ and $\mu$ are Lame's constants, expressed as:
\begin{equation}
    \lambda=\frac{\nu E}{\left(1+\nu\right)\left(1-2\nu\right)}
\end{equation}
\begin{equation}
    \mu=\frac{E}{2\left(1+\nu\right)}
\end{equation}

\begin{figure}[!t]
    \centering
    \includegraphics[width = 0.9\linewidth]{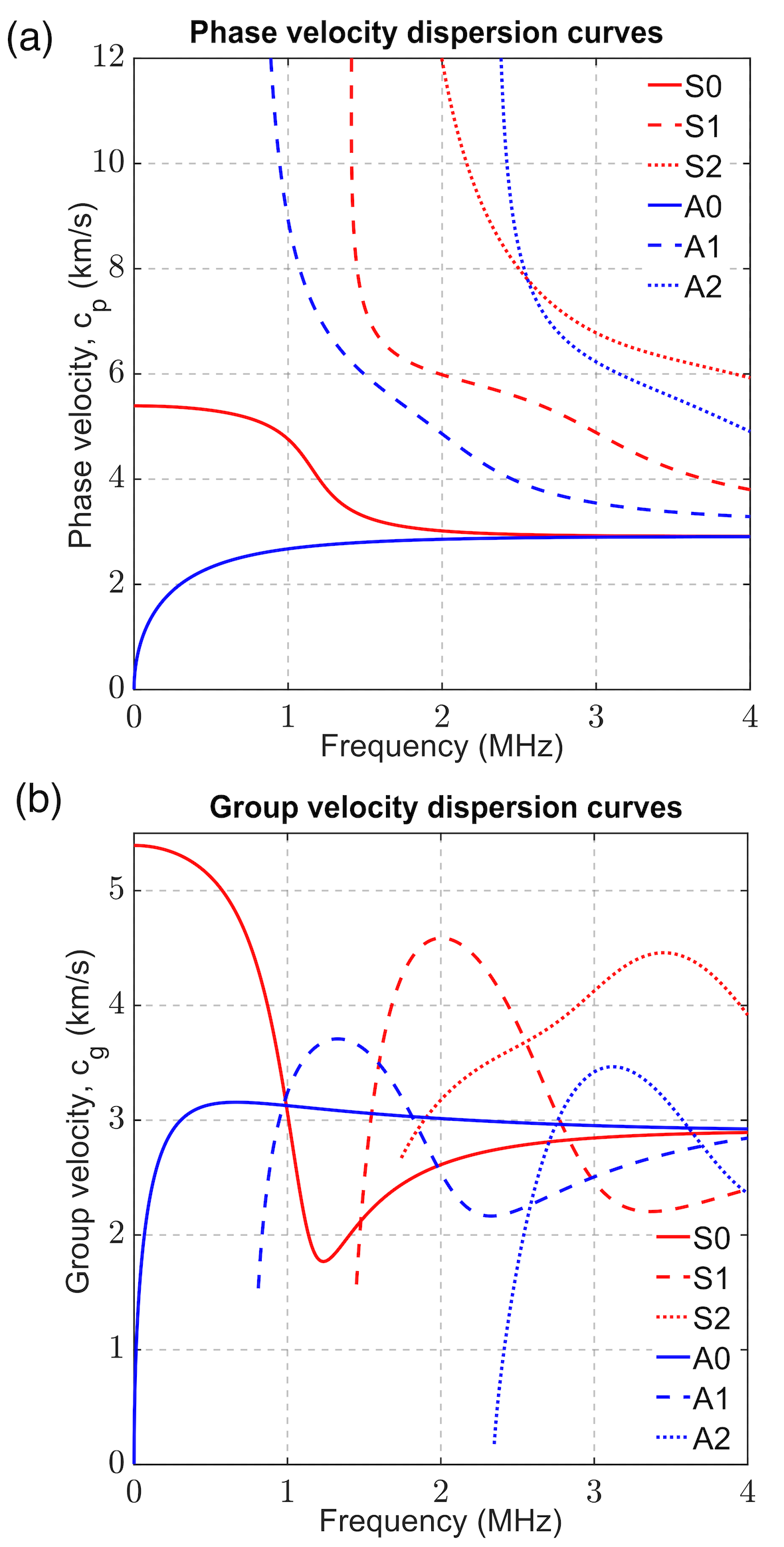}
    \caption{Dispersion curves of (a) phase velocity, and (b) group velocity for various GW modes in \SI{2}{mm} thick Al 6061-T6 plate at \SI{20}{\degree C}.} 
    \label{fig:disp}
\end{figure}

\begin{figure}[!t]
    \centering
    \includegraphics[width = \linewidth]{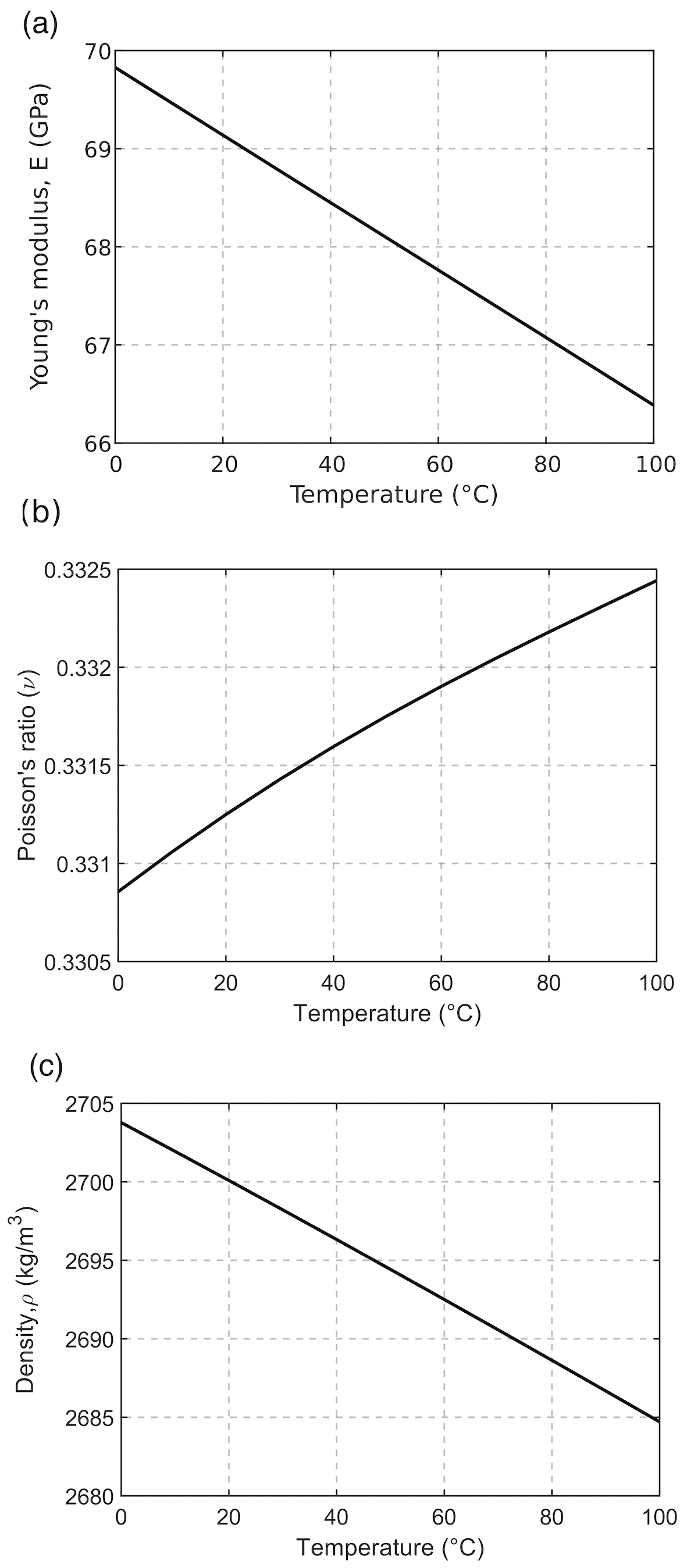}
    \caption{Variation of (a) Young's modulus, (b) Poisson's ratio, and (c) density of Al 6061-T6 with temperature varying from \SI{0}{\degreeCelsius} to \SI{100}{\degreeCelsius}} 
    \label{fig:variation}
\end{figure}

\begin{figure}[!t]
    \centering
    \includegraphics[width = \linewidth]{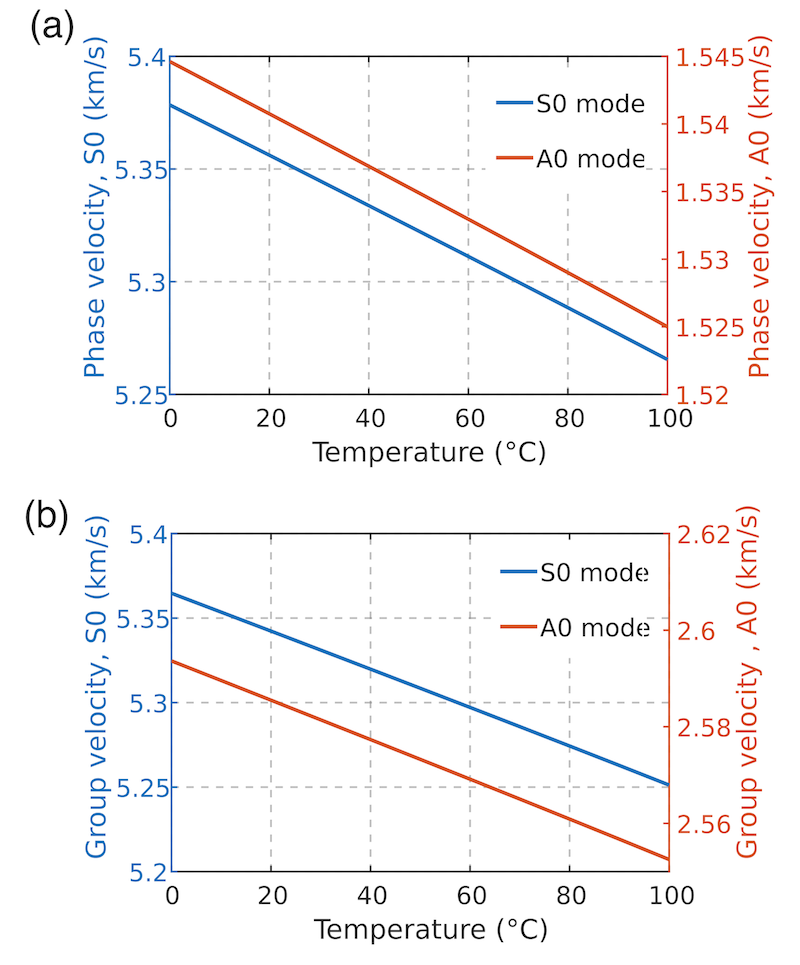}
    \caption{Variation of (a) phase velocity, and (b) group velocity for $S0$ and $A0$ modes with temperature varying from \SI{0}{\degreeCelsius} to \SI{100}{\degreeCelsius}.}
    \label{fig:variationtemp}
\end{figure}

The terms $E$, $\nu$ \& $\rho$ denote the Young's modulus of elasticity, Poisson's ratio and density of the material, respectively.
%
The group velocity $c_g$ of the wave modes can be expressed as \cite{Rose2014}:
\begin{equation}\label{eq:cg}
    c_g=c_p+k \frac{dc_p}{dk}
\end{equation}
Using equations \eqref{eq:sym} and \eqref{eq:asym} and wave number $k$, the phase velocity $c_p$ and group velocity $c_g$ can be calculated for symmetric and anti-symmetric modes in a plate of a specified material ($E,\nu$ and $\rho$), thickness $h$ at frequency $f=\frac{\omega}{2\pi}$. Figure \ref{fig:disp} shows phase velocity and group velocity dispersion curves for a \SI{2}{mm} thick aluminum 6061-T6 alloy plate at room temperature (\SI{20}{\degreeCelsius}).
For our analysis, \SI{150}{\kilo Hz} is chosen as the excitation frequency so that the fundamental symmetric mode ($S0$) and anti-symmetric mode ($A0$) can be easily separated. The variation of material properties of Al 6061-T6 with temperature is shown in Figure \ref{fig:variation}. The material properties are obtained from COMSOL Multiphysics 5.5 material library. The variation of phase velocity and group velocity with temperature ranging from \SI{0}{\degreeCelsius} to \SI{100}{\degreeCelsius} are shown in Figure \ref{fig:variationtemp}(a) and (b), respectively.

\begin{figure}[!t]
    \centering
    \includegraphics[width = 0.95\linewidth]{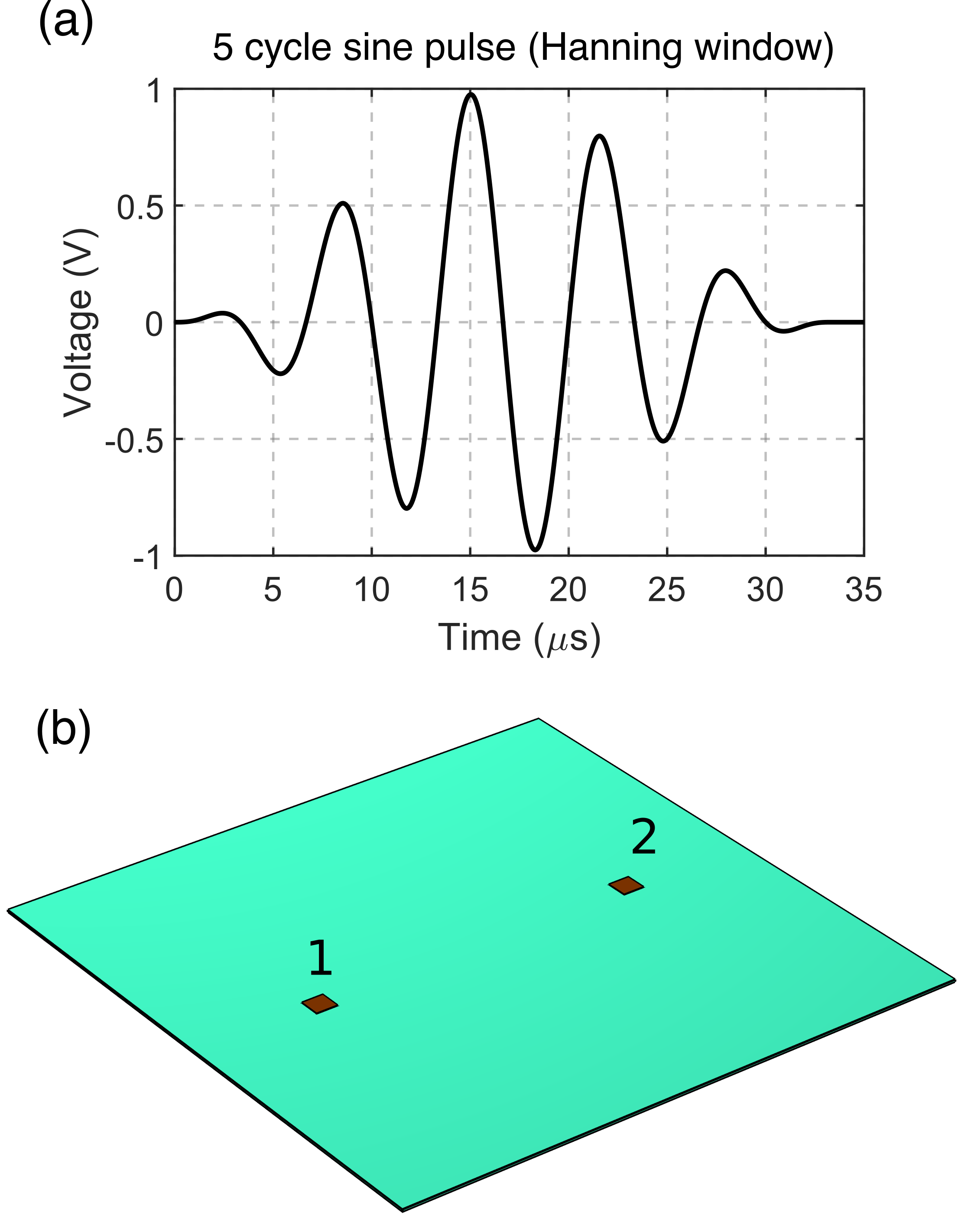}
    \caption{(a) Input signal to transmitter: \SI{150}{\kilo Hz} five-cycle sine pulse shaped by a Hanning window. (b) Setup for studying impact of temperature on group velocity of GW modes: \SI{530}{mm} $\times$ \SI{530}{mm} $\times$ \SI{2}{mm} Al 6061-T6 plate with \SI{20 x 20}{mm} PZT transmitter (1) and receiver (2) spaced apart by \SI{300}{mm} distance. } 
    \label{fig:validationFE}
\end{figure}

The GW propagation in a \SI{530}{mm} $\times$ \SI{530}{mm} $\times$ \SI{2}{mm} Al 6061-T6 alloy plate with surface mounted \SI{20 x 20 x 0.4}{mm} PZT-5H transducers was studied using FE simulations performed with COMSOL Multiphysics 5.5, wherein the isotropic aluminum plate and PZT transducers were modeled using \textit{Structural Mechanics Module} coupled with \textit{Piezoelectric Solid Interaction} physics and analyzed using \textit{Time Implicit Study}. The accuracy of FE simulation of wave propagation is primarily determined by the mesh size and time increment used in the simulation \cite{LECKEY2018187}. For excitation frequency ($f$) of \SI{150}{\kilo Hz}, the wavelengths of the $S0$ and $A0$ modes in the Al plate are approximately \SI{36}{mm} and \SI{17}{mm}, respectively. Previous research on Lamb wave propagation simulations recommends the largest mesh element size to be smaller than $\frac{1}{6}$ of the mode wavelength \cite{lei2019multiphysics}. Thus, the largest and smallest mesh sizes in the Al plate are set to \SI{5}{mm} and \SI{2}{mm} respectively. Physics controlled \textit{Extra Fine} mesh setting was applied for the surface mounted PZT-5H transducers. The time step was chosen to be \SI{0.1}{\micro s} $(>\frac{1}{60 f})$ . The transmitter PZT was actuated using a 5-cycle sine pulse shaped by a Hanning window as shown in Figure \ref{fig:validationFE}(a) expressed as:
\begin{equation}
    V=0.5\left(1-\cos\left(\frac{2\pi f t}{5}\right)\right)\sin(2 \pi f t)
\end{equation}

\begin{figure}[!t]
    \centering
    \includegraphics[width = \linewidth]{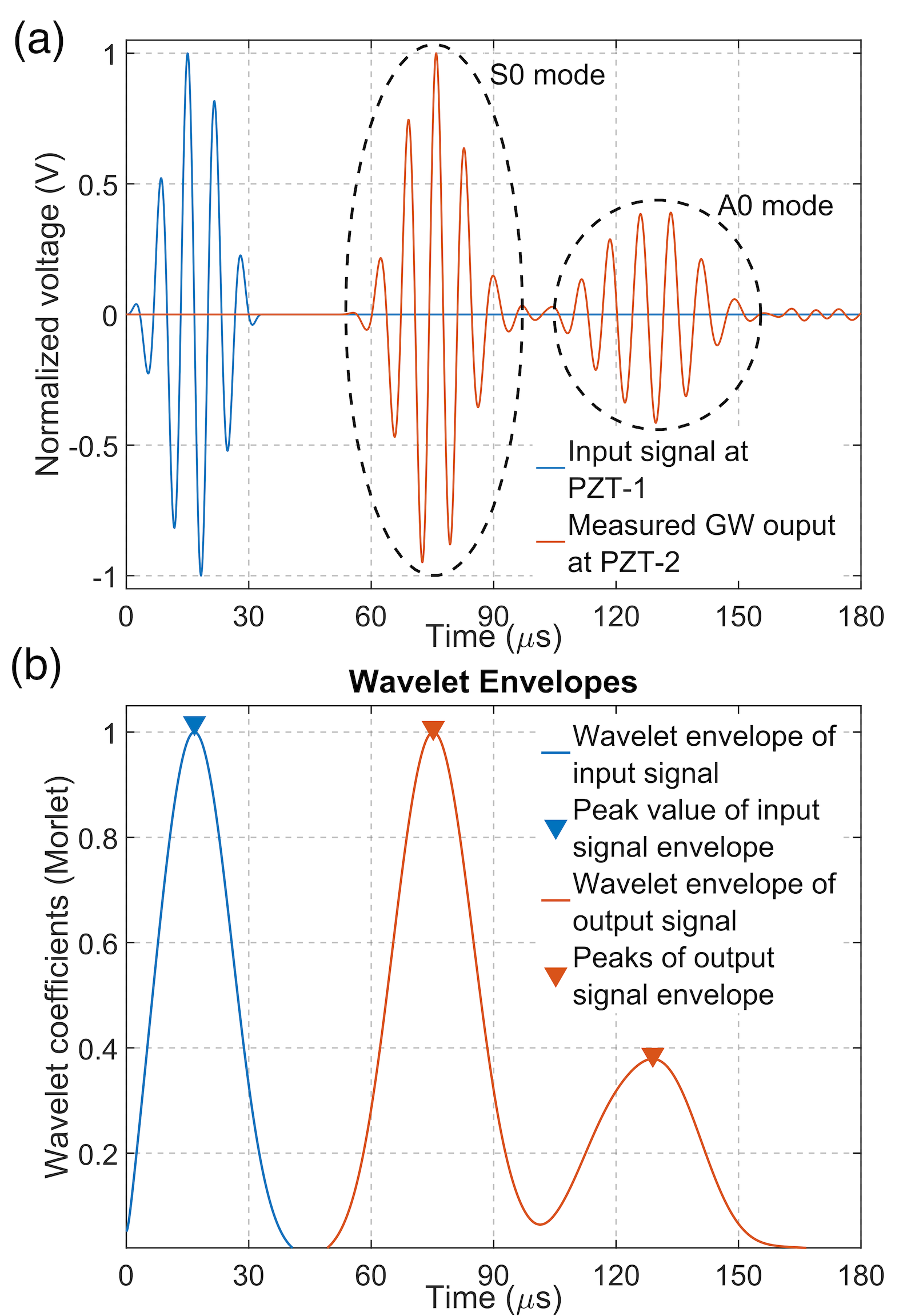}
    \caption{(a) Input signal (blue trace) at transmitter (PZT-1) and GW signal recorded at receiver (PZT-2), showing $S0$ and $A0$ modes. (b) Time of flight for each mode is measured by the peak to peak distance of the wavelet envelopes of the corresponding signal at receiver from the time instant corresponding to peak value of input signal envelope.} 
    \label{fig:wavelet2pzt}
\end{figure}

For validation of the impact of temperature on group velocity of the GW modes obtained from the FE model, a baseline (damage free) Al 6061-T6 alloy plate of dimensions \SI{530}{mm} $\times$ \SI{530}{mm} $\times$ \SI{2}{mm} with two surface mounted PZT-5H transducers as shown in Figure \ref{fig:validationFE}(b) was modeled. The 5-cycle sine pulse shaped by a Hanning window was applied at the transmitter PZT (1) and the output signal was recorded at the receiver (2), separated by a distance of \SI{300}{mm} from the transmitter. \textit{Low reflecting edge} boundary condition was imposed on the four free edges of the plate to avoid reflections of the $S0$ \& $A0$ modes from the edges. The total simulation time was set to \SI{150}{\micro s} with a time step of \SI{0.1}{\micro s}. The total number of elements in the FE model were \SI{155614}{}, with \SI{914717}{} degrees of freedom. The computation time for each simulation obtained on a \SI{64}{}-bit Intel\textregistered~Core\texttrademark-i7-7700HQ \SI{2.80}{GHz} CPU with \SI{16}{GB} RAM workstation was \SI{5700}{\second}.
The \textit{time of flight} for each wave mode was measured using continuous wavelet transform of the output signal measured at the receiver PZT (with Morlet wavelet used as mother wavelet), as shown in Figure \ref{fig:wavelet2pzt}. Table \ref{tab:grpvelocity} shows excellent agreement between the group velocities at \SI{20}{\degree C} obtained from FE simulations at to those obtained from analytical calculations (equation \eqref{eq:cg}).

\begin{table}[!t]
\begin{center}
\small
\caption{Comparison of FE measured group velocities with analytically computed values for $S0$ and $A0$ modes at \SI{20}{\degree C}}

\label{tab:grpvelocity}
\begin{tabular}{|l|l|l|}
\hline 
\textbf{Wave mode}      & \textbf{FEM result  [\SI{}{km/s}]}  & \textbf{Analytical [\SI{}{km/s}]} \\ \hline \hline
S0   & $5.128$ & $5.342$       \\ [0.5ex]
A0  & $2.671$ & $2.586$        \\ \hline
\end{tabular}
\end{center}
\end{table}

\begin{figure}[!t]
    \centering
    \includegraphics[width = 0.85\linewidth]{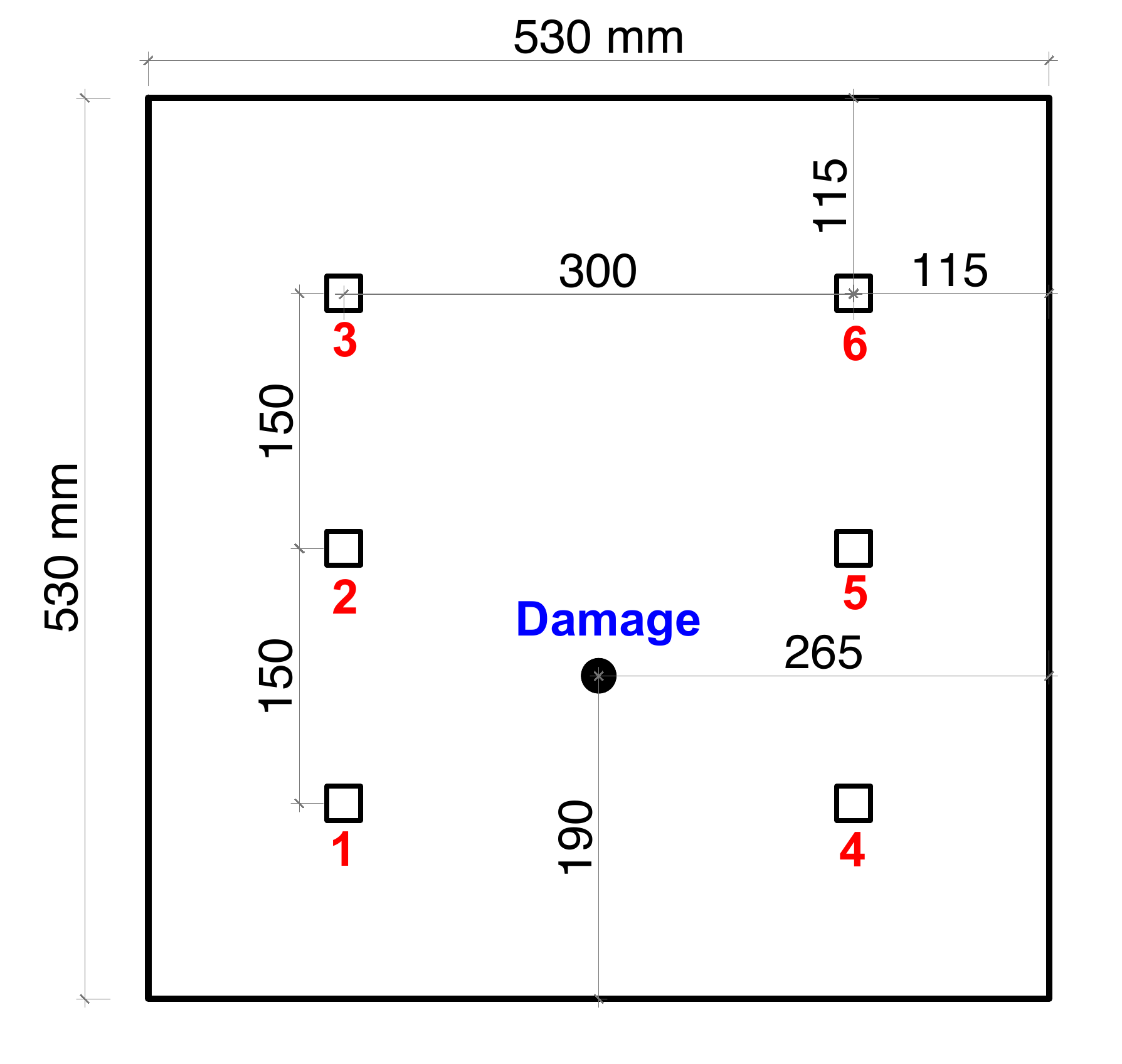}
    \caption{Layout of Al 6061-T6 plate with network of six PZT transducers used for FE simulations. Three types of damages are considered (listed in Table \ref{tab:defectsets}).} 
    \label{fig:6PZTlayout}
\end{figure}

\begin{figure}[!t]
\centering
\includegraphics[width=0.95\linewidth]{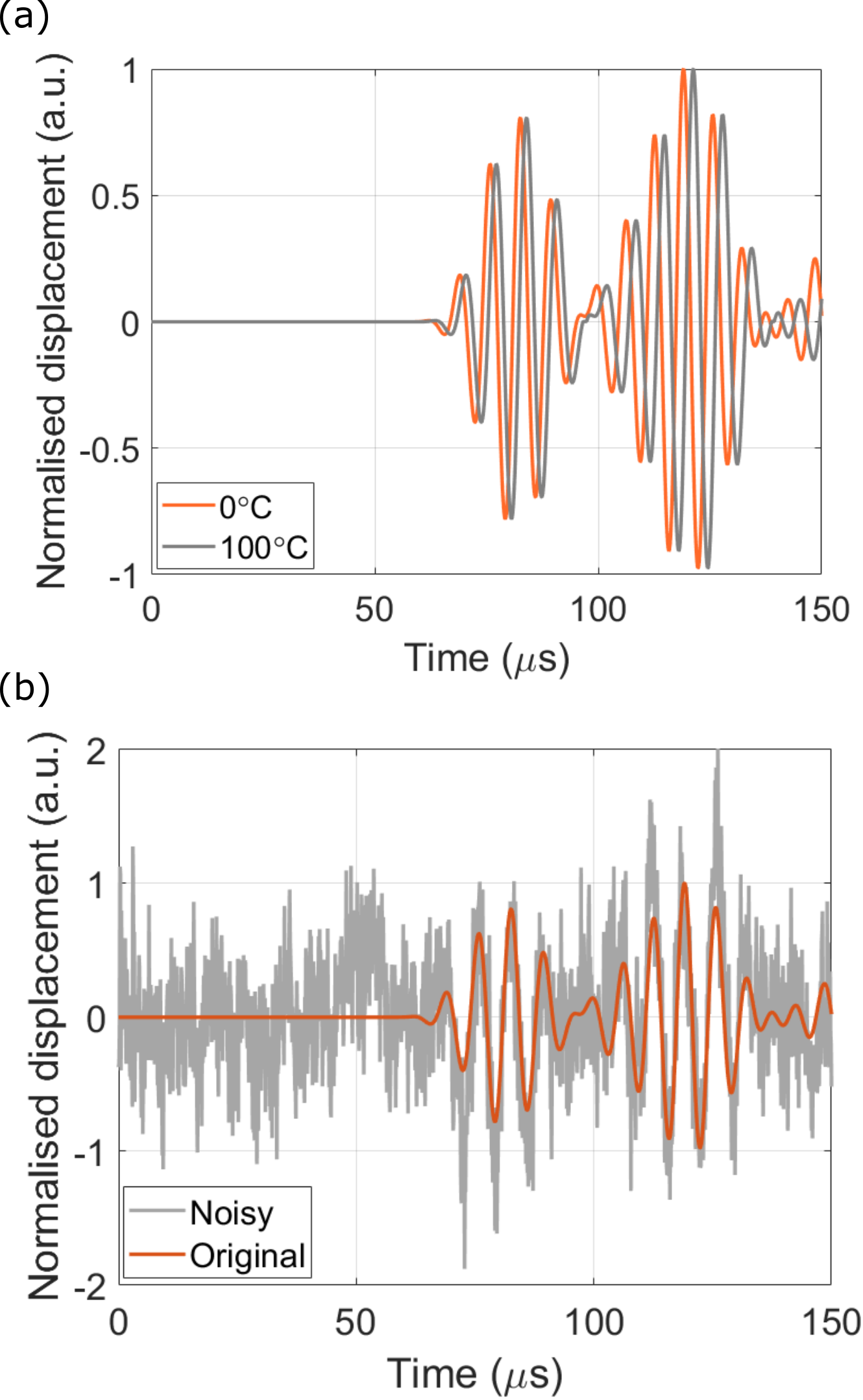}
\caption{(a) Effect of temperature on baseline GW data corresponding to sensor path P15. (b) Signal with added white noise and pink noise with \SI{4.5}{dB} SNR, illustrated for sensor path P15 at \SI{0}{\degreeCelsius}. }
\label{fig:temp_noisy}
\end{figure}

To evaluate the performance of the damage identification and localization method, FE simulations were carried out on an aluminum 6061-T6 alloy plate of dimensions \SI{530}{mm} $\times$ \SI{530}{mm} $\times$ \SI{2}{mm} with $6$ PZT-5H transducers and three types of damages (rivet hole, added mass and notch; parameters listed in Table \ref{tab:defectsets}) introduced in the lower half of the plate, as shown in Figure \ref{fig:6PZTlayout}. Each time-series had a duration of \SI{150}{\micro s}, obtained with a sampling rate of \SI{10}{MHz} for seven separate transmitter-receiver paths (P12, P15, P24, P25, P26, P53 and P56, where P\textit{xy} denotes the time-series output obtained with PZT \textit{x} as transmitter and PZT \textit{y} as receiver). The damage is located at the intersection of the diagonal paths in the lower half of the plate i.e. paths P15 and P24. The FE model had \SI{211069}{} elements and \SI{1210935}{} degrees of freedom.
\begin{table}[!t]
\begin{center}
\centering
\small
\caption{Types and parameters of damages studied for evaluation of the proposed damage classification and localization algorithm. The one-hot encoding based class labels used for CNN classification are also listed.}

\label{tab:defectsets}
\begin{tabular}{|l|l|l|}
\hline
\textbf{Damage}      & \textbf{Parameters} &\textbf{Label} \\ \hline \hline
Baseline & - &$1000$\\
Rivet hole & Diameter \SI{20}{mm} & $0100$\\ 
Added mass & Diameter \SI{23}{mm}, weight \SI{20}{g}  &$0010$\\ 
Notch      & \SI{60}{mm} $\times$ \SI{7}{mm} &$0001$ \\ \hline
\end{tabular}
\end{center}
\end{table}
The computation time for each simulation (for the mentioned seven paths) obtained on a \SI{64}{}-bit Intel\textregistered~Core\texttrademark-i7-7700HQ \SI{2.80}{GHz} CPU with \SI{16}{GB} RAM workstation was \SI{15290}{\second}. Considering the long time required for performing simulations, the following method was used to generate datasets at various temperatures. Group velocities for the symmetric and anti-symmetric modes were obtained from analytical solutions of equations \eqref{eq:sym} and \eqref{eq:asym}, using the expression in equation \eqref{eq:cg}. FE simulations were performed for healthy (damage-free) and various damage conditions at \SI{0}{\degreeCelsius} temperature. The temperature-dependent elastic properties (Figure \ref{fig:variation}) were used to determine group velocities of $S0$ \& $A0$ modes for temperatures ranging from \SI{0}{\degreeCelsius} to \SI{100}{\degreeCelsius}.
Generally, both $S0$ and $A0$ modes are commonly used for damage detection in isotropic plates \cite{SU2006753}. However, $S0$ mode is faster and has much lower attenuation than $A0$ mode and therefore can propagate longer distances \cite{Masurkar2017,SU2006753}. In our present study, we focused on interaction of $S0$ mode with the damage. For estimating the signal at temperature $T$, the GW signal obtained from FE simulation for \SI{0}{\degreeCelsius} for any transmitter-receiver path on the panel was time-shifted by an amount $\Delta t_T$:
\begin{equation}
    \Delta t_T=\frac{d}{c_{gT}}-\frac{d}{c_{g0}}
\end{equation}
where $d$ is the distance between the transmitter-receiver pair under consideration, $c_{gT}$ is the analytically calculated group velocity of $S0$ mode at temperature $T$, and $c_{g0}$ is the analytically calculated group velocity of $S0$ mode at \SI{0}{\degreeCelsius}. 
The variation of signal amplitude with temperature was computed using the amplitude of $S0$ wave packet for temperature ranging from \SI{0}{\degreeCelsius} to \SI{100}{\degreeCelsius} obtained analytically using \textit{Dispersion calculator, DC} software \cite{dlrsoftware}. The maximum amplitude value of $S0$ wave packet was observed to vary linearly with increasing temperature. 
An analytical relation between $A_f$ (ratio of maximum amplitude of $S0$ wave packet at temperature $T$ to that at \SI{0}{\degreeCelsius}) and temperature $T$ (specified in units of \SI{}{\degreeCelsius}) ranging from \SI{0}{\degreeCelsius} to \SI{100}{\degreeCelsius} is obtained using MATLAB \textit{Curve Fitting Toolbox}:

\begin{equation}
    A_f=1 + \frac{T~[\SI{}{\degreeCelsius}]}{3550}
    \label{eq:AmpT}
\end{equation}
The GW signal $s_{T}(t)$ at temperature $T$ can thus be obtained:
\begin{equation}\label{eq:sT}
    s_{T}(t)=A_{f} s_{0}(t+\Delta t_T)
\end{equation}
where $s_{0}(t)$ is the signal at \SI{0}{\degreeCelsius}
%
%
This technique helps in reducing the computational time required for obtaining GW signals for all sensor paths at various temperatures, as this only requires time-consuming FE simulations at \SI{0}{\degreeCelsius} temperature. The data obtained at \SI{0}{\degreeCelsius} was used to generate data at other temperatures obtained using equation \eqref{eq:sT}. To verify the accuracy of this method, we evaluated the cross-correlation coefficient between the signal obtained using equation \eqref{eq:sT} and that obtained from FE simulations performed at \SI{50}{\degreeCelsius} and \SI{100}{\degreeCelsius}. The cross-correlation coefficient was observed to be $>$\SI{99}{\percent}, and hence this method is justified to reduce computational complexity.
Representative temperature-adjusted time-series for path P15 at two different temperatures: \SI{0}{\degreeCelsius} and \SI{100}{\degreeCelsius}, are shown in Figure \ref{fig:temp_noisy}(a).
To incorporate uncertainties caused by variation in other EOCs and experimental uncertainties, we also added white noise and pink noise to the data:

\begin{equation}
s_{n,T}(t) = s_T(t)+ \beta\max(s_T(t))\left[w(t)+p(t)\right]
\label{EQ:Noise}
\end{equation}
where $s_{n,T}(t)$ is the signal at temperature $T$ with added noise, $\max(s_T(t))$ is the peak amplitude of signal $s_T(t)$, $\beta$ is the noise scaling factor, and $w(n), p(n)$ are white noise and pink noise respectively, generated using normal distribution with zero mean and standard deviation of $0.1$. We studied the performance of the damage assessment method for three different signal-to-noise-ratio (SNR) values: \SI{2}{dB}, \SI{4.5}{dB} and \SI{10}{dB}. Figure~\ref{fig:temp_noisy}(b) shows signal with added noise for $SNR =$ \SI{2}{dB} for sensor path P15 at \SI{0}{\degreeCelsius}. 

\section{Simulation results and discussion}

\begin{figure}[!t]
\centering
\includegraphics[width=\linewidth]{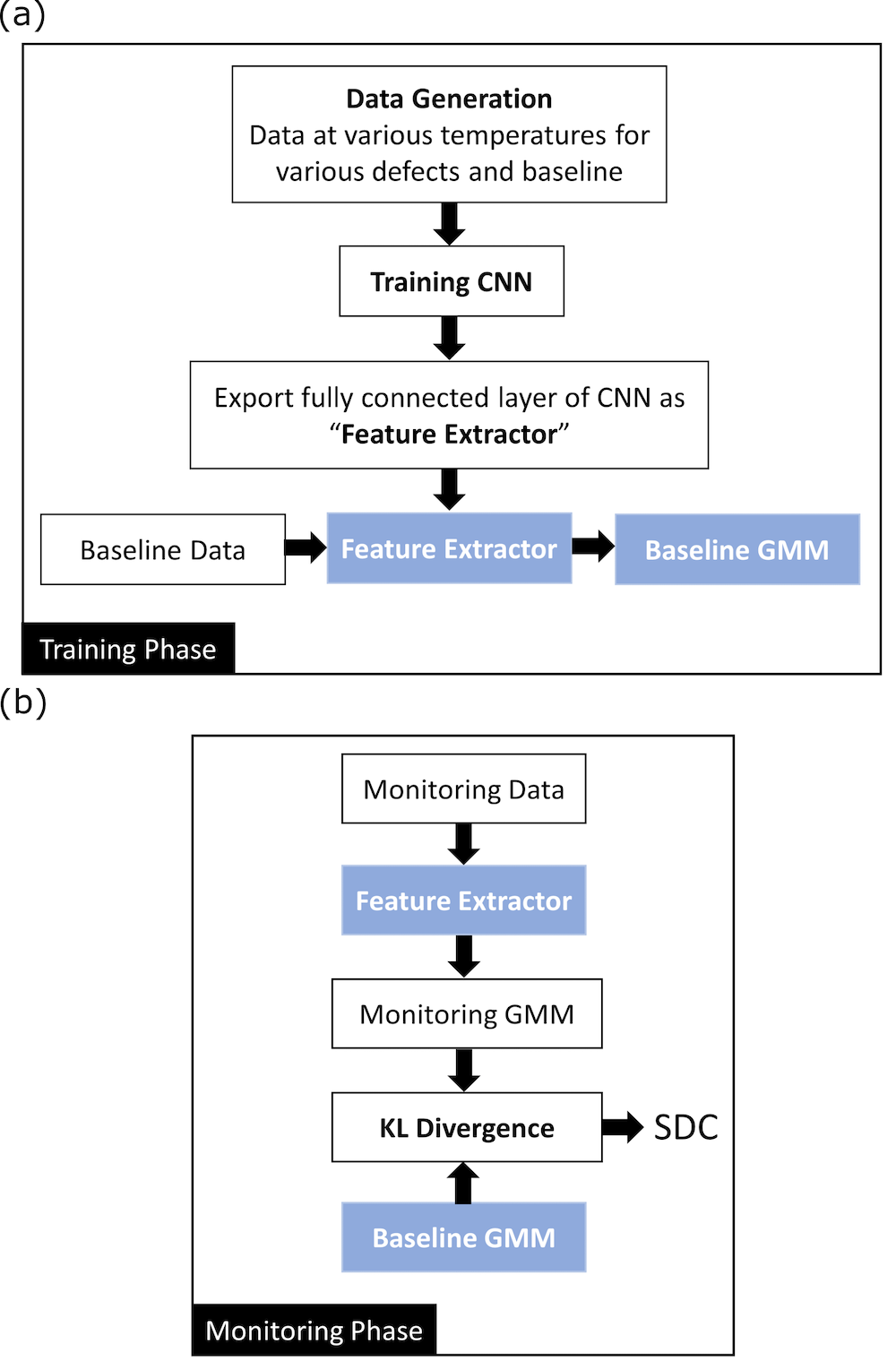}
\caption{Illustration of steps involved in training and monitoring phases. (a) The training phase involves data generation for training CNNs. It also involves modeling the baseline GMM. (b) Monitoring phase involves feature extraction from monitoring data. The feature extractor and baseline GMM are imported from training phase for computing SDC.}
\label{fig:2phases}
\end{figure}

\begin{figure}[!t]
\centering
\includegraphics[width=0.8\linewidth]{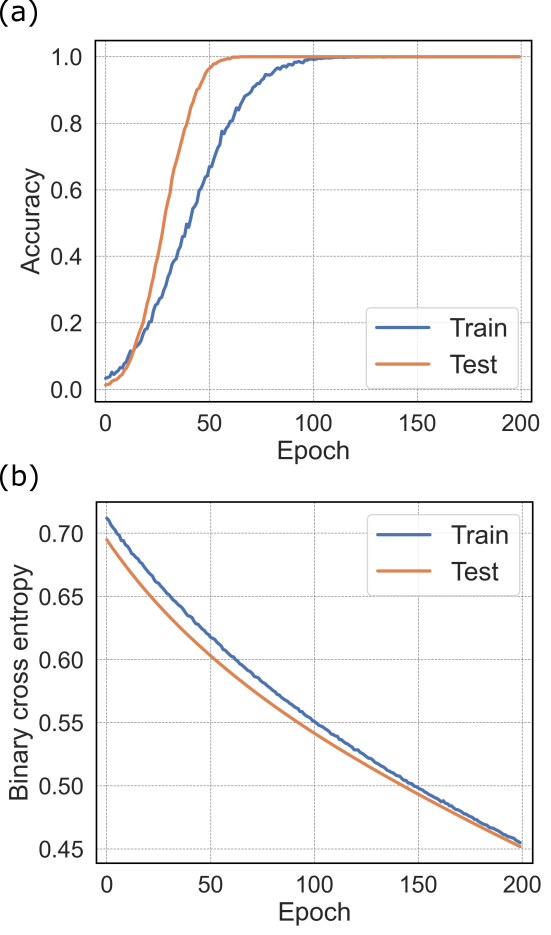}
\caption{(a) Accuracy, and (b) loss curves for training and testing of 1D CNN corresponding to sensor path P15 for $200$ epochs.}
\label{fig:LossCurve}
\end{figure}

\begin{figure*}[!t]
\centering
\includegraphics[width=0.8\linewidth]{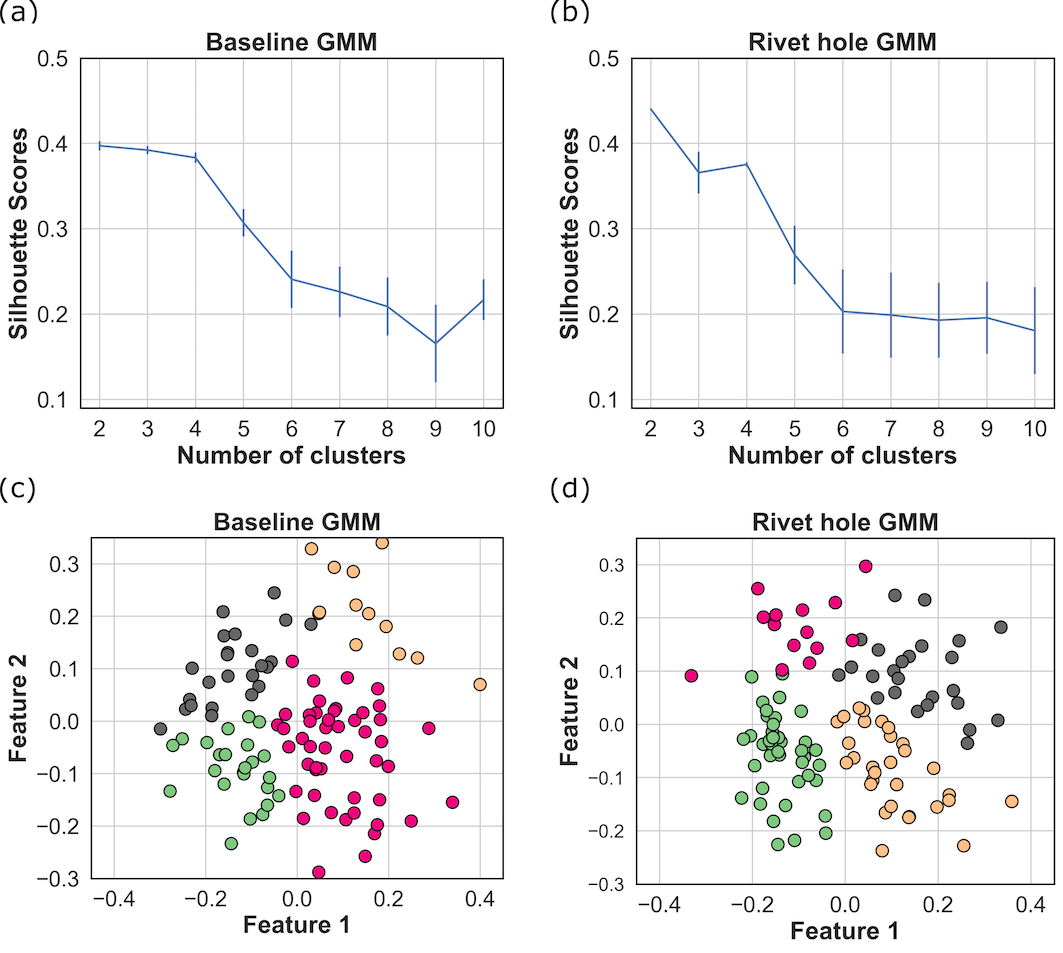}
\caption{For GMM-based clustering, optimum number of clusters are selected using silhouette score. Silhouette scores for various number of clusters for path P15 are shown for (a) baseline data, and (b) rivet hole data. Distributions formed using optimum number of clusters $(4)$ are shown for two prominent features for (c) baseline data, and (d) rivet hole data. The two most prominent features were determined out of $16$ features computed by the feature extractor layer in the CNN using PCA.}
\label{fig:GMM}
\end{figure*}

\begin{figure*}[!t]
\centering
\includegraphics[width=\linewidth]{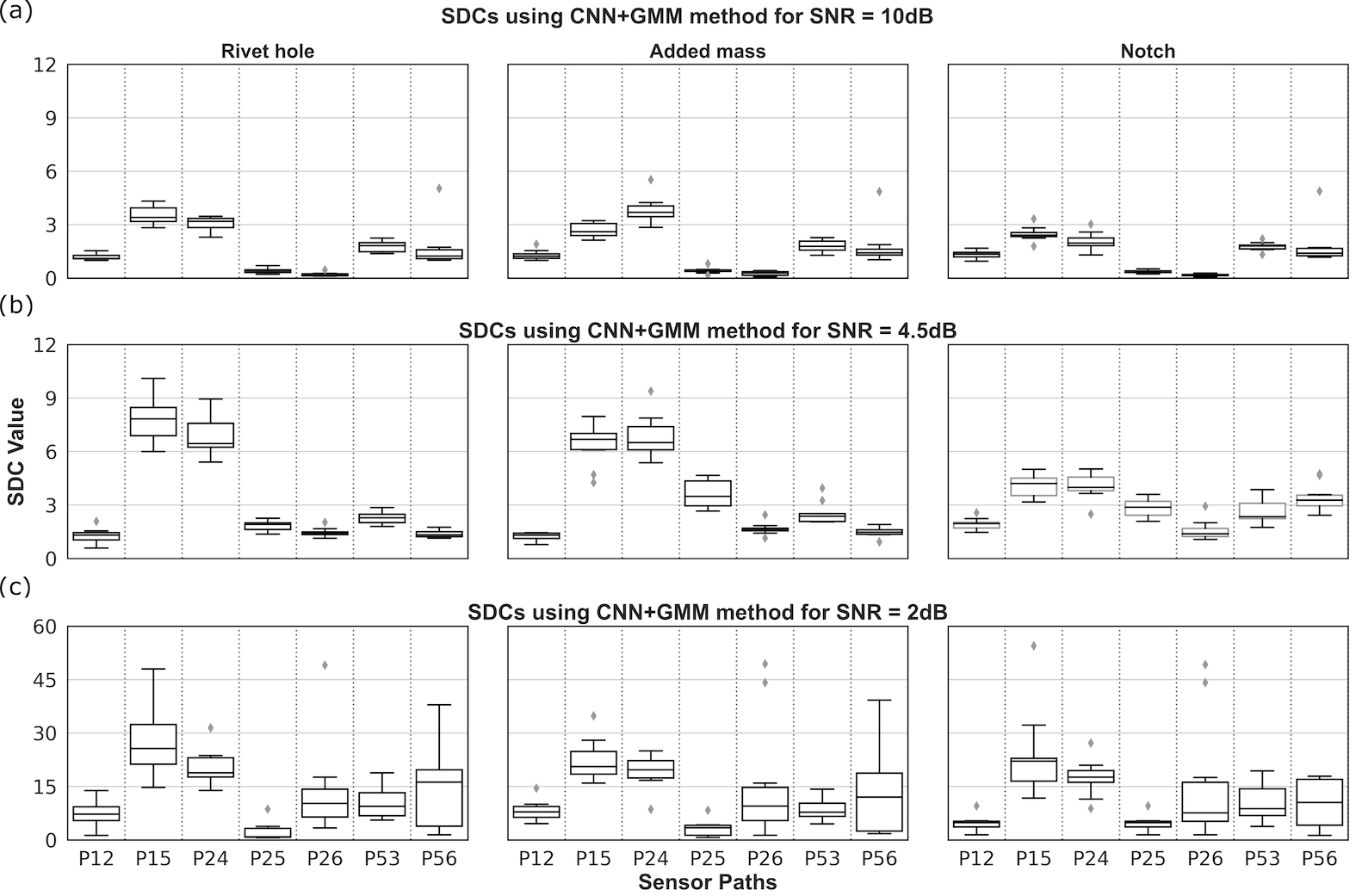}
\caption{Trends in SDCs computed using equation~\eqref{eq:KL} as described in CNN+GMM method.
For each of seven sensor paths, $10$ values of SDCs are computed for $10$ test cases. Three different defects are shown in $3$ columns (rivet hole, added mass and notch). Three different signal to noise ratio (SNR) levels as shown in: (a) \SI{10}{dB}, (b) \SI{4.5}{dB}, and (c) \SI{2}{dB}.}
\label{fig:SDC_3x3}
\end{figure*} 

\begin{figure*}[!t]
\centering
\includegraphics[width=0.9\linewidth]{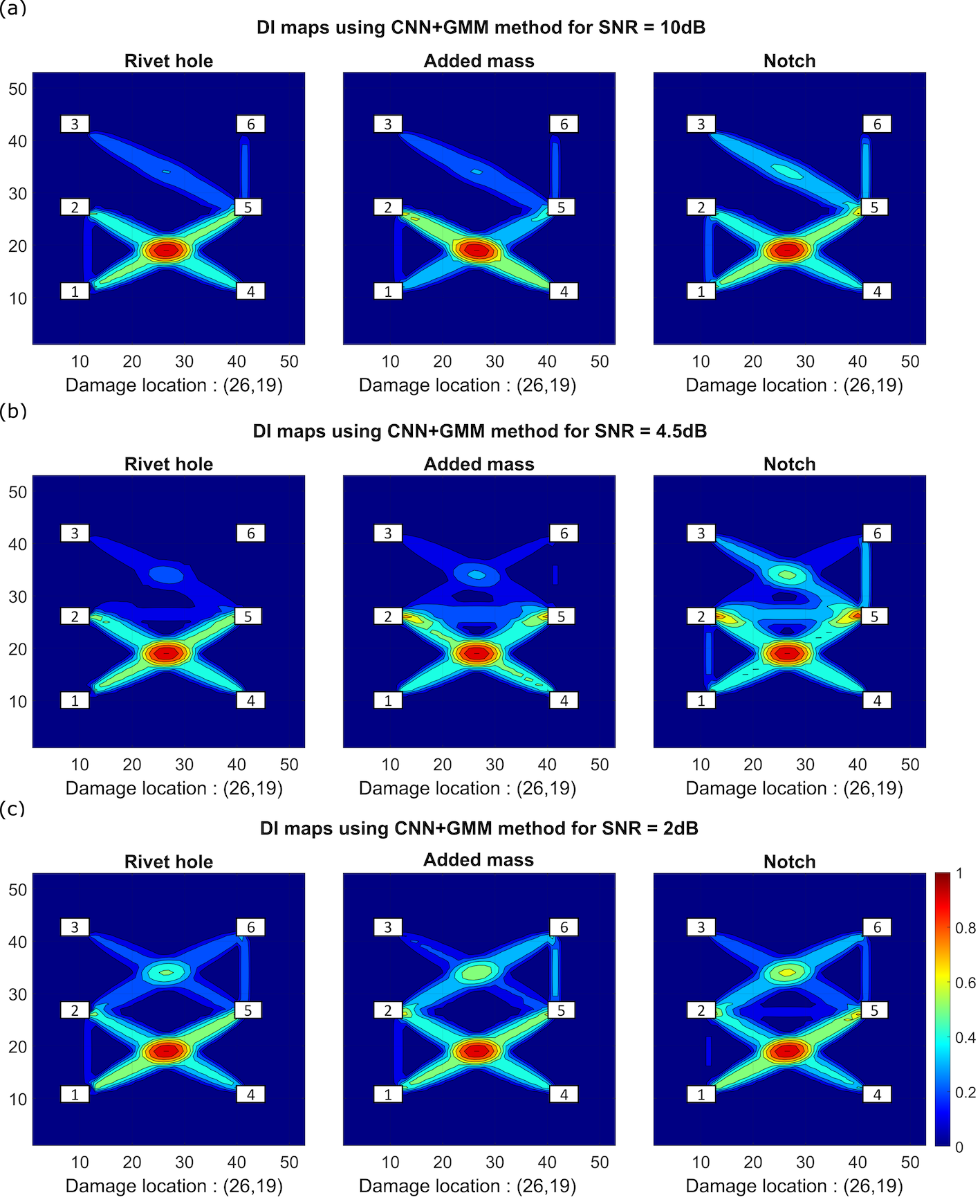}
\caption{DI maps obtained using SDC based damage localization algorithm. Mean value of SDCs across $10$ test cases shown in Figure \ref{fig:SDC_3x3} was used to represent SDC of each sensor path. Three different defects are shown in $3$ columns (rivet hole, added mass and notch). Three different signal to noise ratio (SNR) levels as shown in: (a) \SI{10}{dB}, (b) \SI{4.5}{dB}, and (c) \SI{2}{dB}.}
\label{fig:DI_3x3}
\end{figure*}

To demonstrate and study the performance of the temperature compensated damage localization technique, the implementation was carried out in two phases: training phase and monitoring phase, as shown in Figure \ref{fig:2phases}. The training phase involves data generation and training the CNNs (one CNN per path). The temperature affected GW data set with added noise described in the previous section was used for this purpose. The CNNs were trained using data corresponding to healthy i.e. undamaged (baseline) condition, as well as damaged conditions (rivet hole, added mass and notch). The penultimate layer of the trained CNN was then imported as feature extractor. Using the feature extractor, features for baseline data were computed and used for creating baseline GMM. These two modules, feature extractor and baseline GMM were imported as-is in the next phase i.e. monitoring phase. The monitoring phase was designed such that it is readily deployable for real-time SHM. In this phase, GW data is collected from the structure being monitored. Using the feature extractor imported from training phase, features were computed for monitoring data which were then used to form monitoring GMM. Finally, the KL divergence between baseline GMM and monitoring GMM was computed. The values of KL divergence computed for each path are used as SDC for damage localization algorithm.

\subsection{Performance analysis of CNN+GMM based method for temperature compensated damage localization}

The time-series dataset constructed using method summarised in equation \eqref{EQ:Noise} contained $200 \times 11 \times 4 \times 3 =$ \SI{26400} time-series for each path (\SI{200}{} series generated for each of the \SI{11}{} different temperature values: \SI{0}{\degreeCelsius}, \SI{10}{\degreeCelsius}, \SI{20}{\degreeCelsius} $\dotsc$ \SI{100}{\degreeCelsius}, for each of the \SI{4}{} classes: baseline, rivet hole, added mass, notch) and \SI{3}{} different SNR values: \SI{2}{dB}, \SI{4.5}{dB} and \SI{10}{dB}). Out of these \SI{26400}{} time-series, \SI{13200}{} were used for training the CNNs and the remaining series were divided into \SI{10}{} sets that were used for performance analysis in monitoring phase. 
The training dataset (\SI{4400}{} series for each SNR for each path) was split in the proportion $75:20:5~(\%)$ resulting in \SI{3300}{} for training, \SI{88}{} for testing, and \SI{220}{} for validation of the CNNs. Each class was associated with a label generated with one-hot encoding, as listed in Table \ref{tab:defectsets}.
For each 1D CNN, learning rate of \SI{1e-7}{} and batch size of $64$ were used. Binary cross-entropy loss was used as the cost function and Adam optimization scheme was used to enable momentum and adaptive learning rate in the training process. The networks were trained for $200$ epochs. The CNNs were trained on a \SI{64}{}-bit Intel\textregistered~Core\texttrademark-i7-10750H \SI{2.60}{GHz} CPU workstation with \SI{16}{GB} RAM. The models were developed in Python using TensorFlow library and Keras environment.
The architecture of each 1D CNN is shown in Table \ref{tab:table_sim}.


\begin{table}[!tbp]
\caption{Summary of CNN architecture for binary classification of FE generated temperature and noise affected data}
\label{tab:table_sim}
\small
\begin{tabular}{p{2.75cm} | p{1.75cm} | p{1.75cm}}
\hline

\textbf{Layer (type)} & \textbf{Output shape} & \textbf{Number of parameters} \\ \hline

Conv1D & $1499 \times 16$  & $64$ \\
MaxPooling1D & $749 \times 16$ & $0$ \\
Dropout & $749 \times 16$ & $0$ \\
Flatten & $11984$ & $0$  \\
Dense (ReLU) & $16$ & $191760$ \\
Dense (Sigmoid) & $4$ & $68$ \\ \hline
\multicolumn{2}{c|}{\textbf{Total trainable parameters}} & $191892$ \\ \hline
                                 
\end{tabular}
\end{table}

\begin{figure}[!t]
\centering
\includegraphics[width=\linewidth]{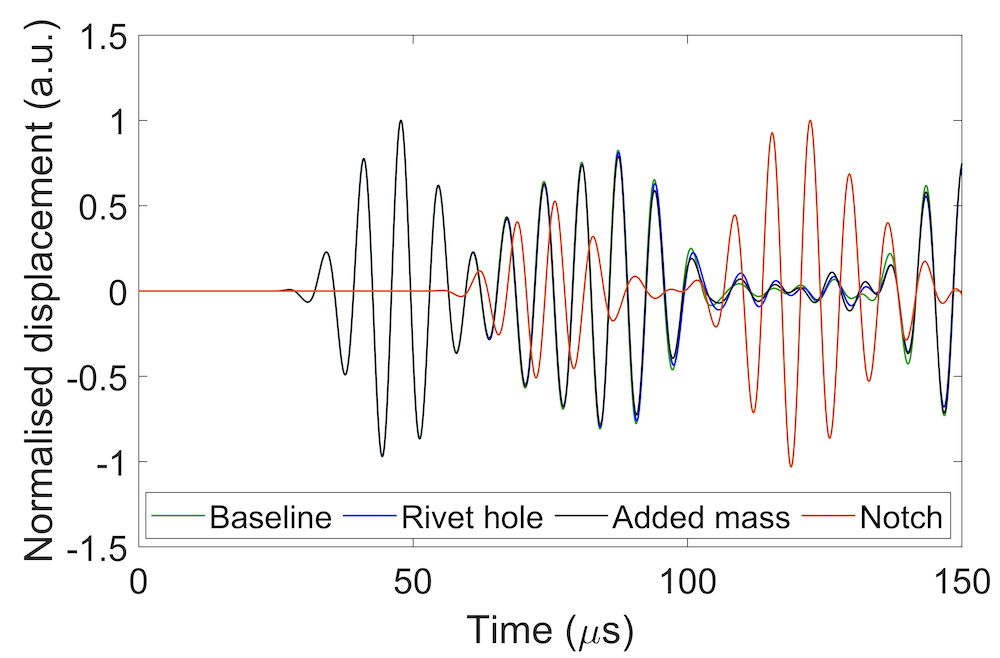}
\caption{GW signal corresponding to baseline, rivet hole, added mass and notch for sensor path P56 at \SI{0}{\degreeCelsius}. No significant difference from baseline is noticed for rivet hole and added mass cases. The signal obtained for notch damage shows significant difference as compared to the other signals.}
\label{fig:EdgeReflections}
\end{figure}

Representative accuracy and loss curves (for $200$ epochs for path P15) are shown in Figure~\ref{fig:LossCurve}. The training loss as well as testing loss decrease smoothly with increasing number of epochs, thus indicating that the model is not over-fitting to the data. For over-fit models, the training error is extremely low while testing error increases sharply as the model tries to over-fit to noise present in the training samples. Perfect classification accuracy $(=1)$ was observed, which is not surprising for a homogeneous material such as aluminum. The penultimate layer of the CNN, used as feature extractor, yields \SI{16}{} features for each time-series consisting of \SI{1500}{} samples. These features were then processed by GMMs. Firstly, a baseline GMM for each path was created using features for \SI{110}{} time-series chosen at random across the \SI{11}{} different temperatures for each SNR. 
In monitoring phase, we evaluated the performance of the method using \SI{10}{} different test cases using time-series data that were not used for training. Each of these \SI{10}{} test cases also contained \SI{110}{} time-series. Using the feature extractor layer, we obtained feature set of size $16 \times 110$ that was used for creating monitoring GMM. 
To determine the optimum number of clusters in the GMM, silhouette score was computed by varying number of clusters for baseline features and monitoring (rivet hole) features as illustrated for path P15 in Figure \ref{fig:GMM}(a) and (b), respectively. In both cases, the value of silhouette score dropped after beyond $4$ clusters, therefore, for both GMMs, four clusters were formed as shown in in Figure \ref{fig:GMM}(c) and (d).
Next, KL divergence value was computed between baseline GMM and monitoring GMM using equation \eqref{eq:KL}. The statistical distribution of values of KL-divergence i.e. SDC for $10$ test cases and seven sensor paths are shown in Figure \ref{fig:SDC_3x3} for the three different SNR values. Both sensor paths containing the defect in-line (P15 and P24) show higher values of KL divergence compared to other five sensor paths which do not contain the defect. The corresponding DI maps are shown in Figure \ref{fig:DI_3x3} for all SNR values. The DI maps are constructed using the mean value of SDC (from \SI{10}{} test cases) for each path.
The distinction between SDC values for paths containing the damage (P15 and P24) and paths without damage is better observed for higher SNRs, however the DI maps accurately highlight the location of the damage even for $SNR=$\SI{2}{dB}. Note that \SI{2}{dB} SNR is unusually low in conventional GW SHM systems, however the proposed method is effective even with such low SNR. 
We observe that the distinction in SDC values for damaged and undamaged paths was more apparent for rivet hole and added mass, as compared to notch defect. This was attributed to the edge reflections in undamaged paths caused by the notch (dimension \SI{60}{mm} $\times$ \SI{7}{mm}). As shown in Figure \ref{fig:EdgeReflections}, FE simulated GW signals for path P56 (that does not contain the damage) corresponding to baseline, rivet hole and added mass are very similar to each other (overlapping curves), whereas significant phase shift was observed for notch.

\begin{figure}[!t]
\centering
\includegraphics[width=0.8\linewidth]{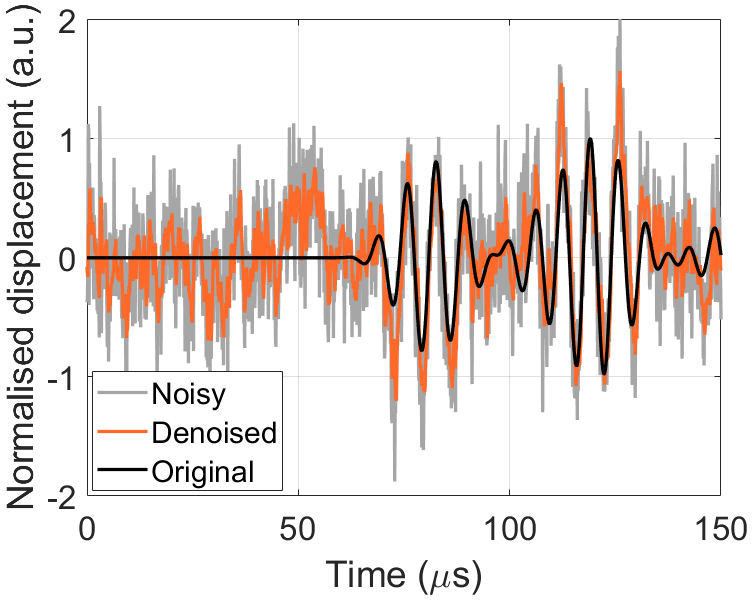}
\caption{Denoising of noisy signal using moving average filter with kernel size $=3$ for GW signal corresponding to sensor path P15 at \SI{0}{\degreeCelsius}.}
\label{fig:denoised}
\end{figure}

\begin{figure*}[!t]
\centering
\includegraphics[width=\linewidth]{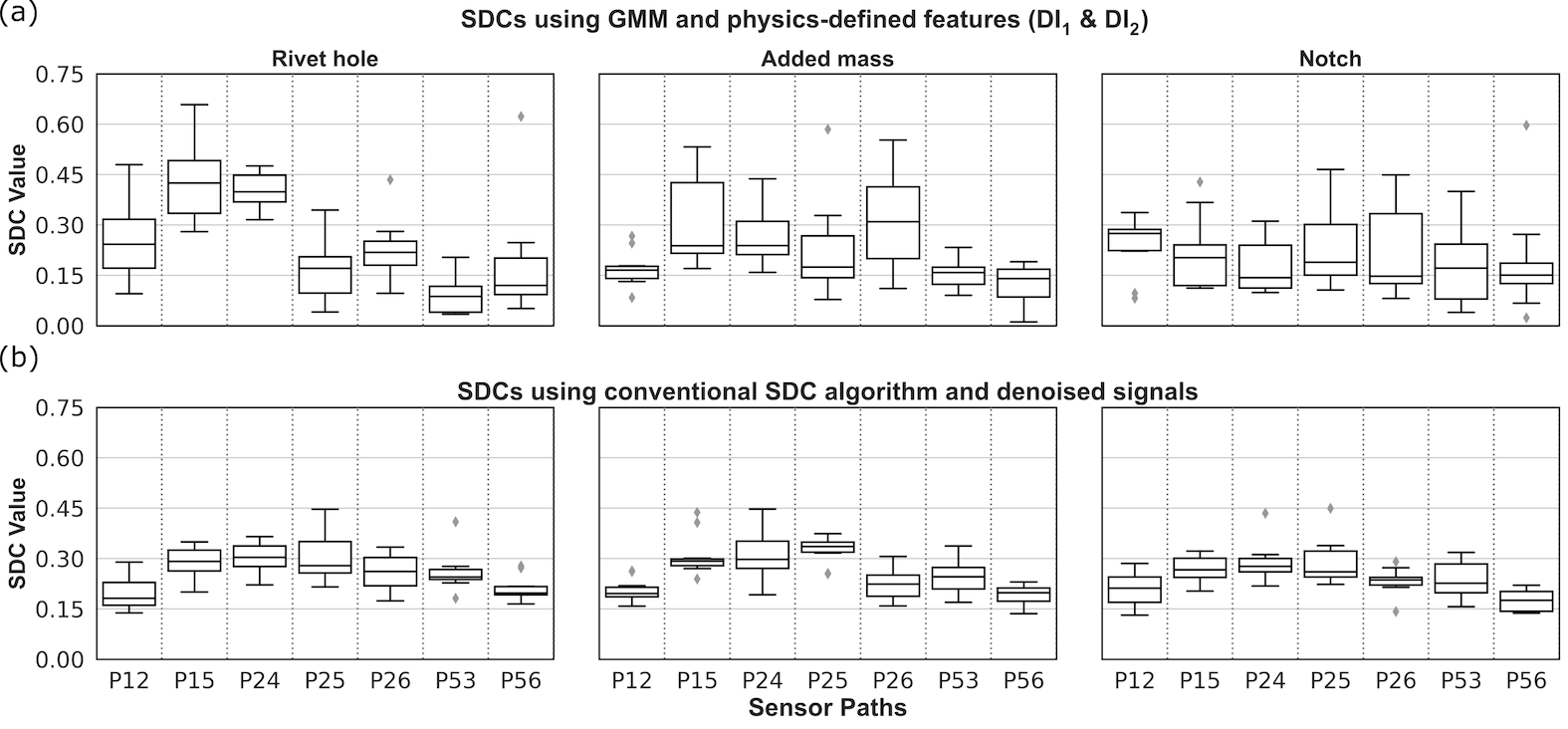}
\caption{Trends in SDCs computed using manually computed features. For each of seven sensor paths, $10$ values of SDCs are computed for $10$ test cases for SNR = \SI{4.5}{dB}. Three different defects are shown in $3$ columns (rivet hole, added mass and notch). SDCs are computed using computed using (a) KL divergence of GMMs modeled using physics-defined features $DI_1$ and $DI_2$ (equations \eqref{EQ:DI1} and \eqref{EQ:DI2}), and (b) conventional SDC definition using CCD metric (equation \eqref{EQ:CCD}) applied to signals denoised using moving average kernel of size $3$.}
\label{fig:SDC_2x3}
\end{figure*}

\begin{figure*}[!t]
\centering
\includegraphics[width=0.9\linewidth]{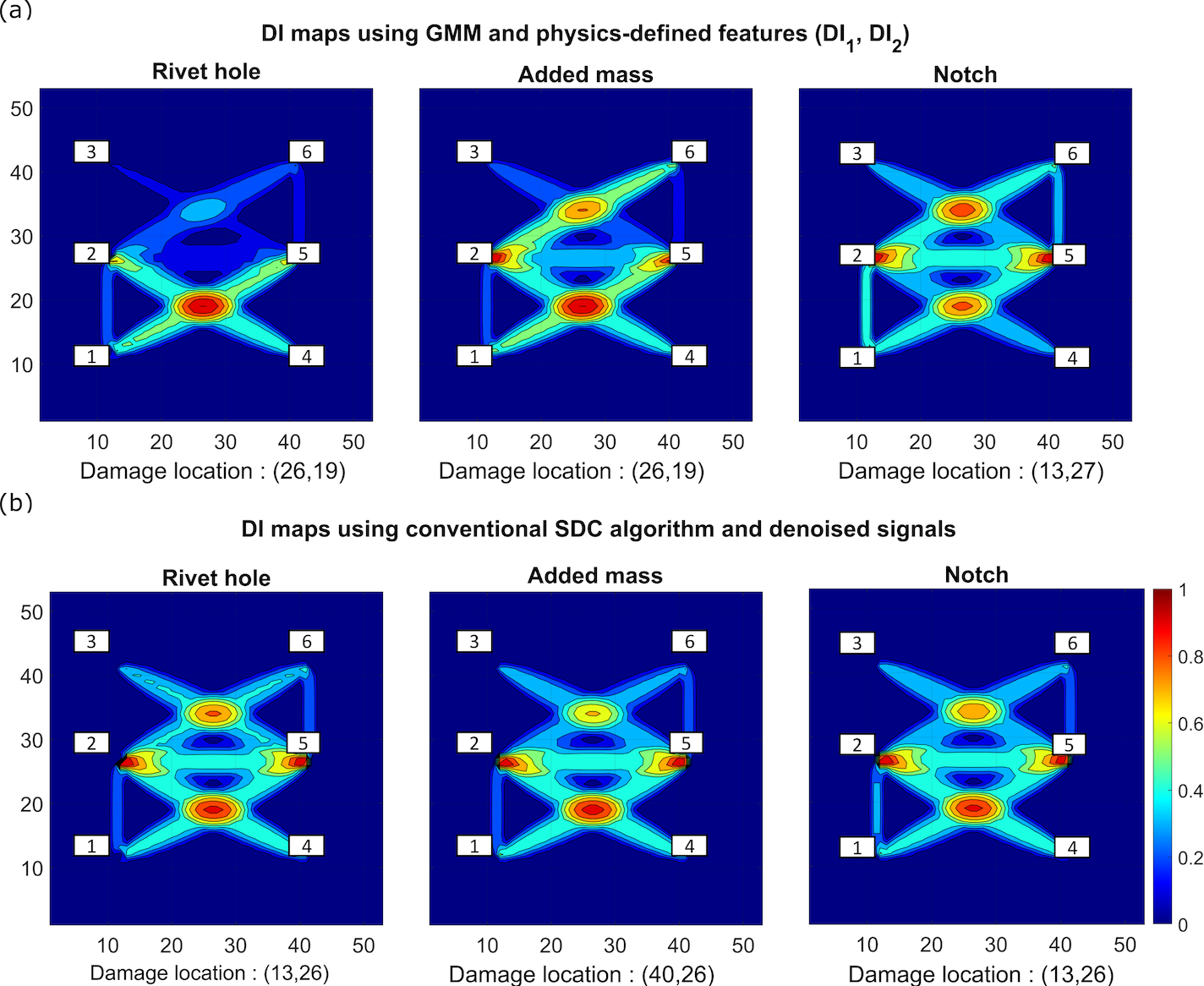}
\caption{
DI maps obtained using SDC based damage localization algorithm. Mean value of SDCs across $10$ test cases shown in Figure \ref{fig:SDC_2x3} was used to represent SDC of each sensor path. Three different defects are shown in $3$ columns (rivet hole, added mass and notch). SDCs are computed using computed using (a) KL divergence of GMMs modeled using physics-defined features $DI_1$ and $DI_2$ (equations \eqref{EQ:DI1} and \eqref{EQ:DI2}), and (b) conventional SDC definition using CCD metric (equation \eqref{EQ:CCD}) applied to signals denoised using moving average kernel of size $3$.
}
\label{fig:DI_2x3}
\end{figure*}

\subsection{Comparison of CNN+GMM method vs. GMM using physics-defined features}
\label{sec:DI1&DI2}
To evaluate the efficacy of the CNN+GMM method, we compared the performance reported above to the method proposed by Ren et al. \cite{ren2019gaussian}, that uses 2 features ($DI_1$ and $DI_2$) and GMM for EOC compensation: 
\begin{equation}
DI_1 = 1 - \sqrt{\frac{\{\int_{t_0}^{t_1}S_b(t)S_m(t)dt\}^2}{\{\int_{t_0}^{t_1}{S_b(t)}^2dt\int_{t_0}^{t_1}{S_m(t)}^2dt\}}}
\label{EQ:DI1}
\end{equation}

\begin{equation}
DI_2 = \sqrt{\frac{\int_{f_0}^{f_1}\{|S_b(f)| - |S_m(f)|\}^2df}{\int_{f_0}^{f_1}|S_b(f)|^2df \int_{f_0}^{f_1}|S_m(f)|^2df}}
\label{EQ:DI2}
\end{equation}

Feature $DI_1$ quantifies the correlation coefficient deviation between baseline signal $S_b(t)$ and monitoring signal $S_m(t)$, whereas feature $DI_2$ quantifies the difference in the Fourier spectra of baseline ($S_b(f)$) and monitoring ($S_m(f)$) signals, normalized to their respective energies.
Here, $t_0$ is time of arrival for the $S0$ mode and $t_1 = t_0 + t_w$ is the exit time, where $t_w=$ \SI{33.3}{\micro s} is pulse width of actuation pulse (5-cycle sine wave with Hanning window in this case). Table \ref{tab:TAO_manual} lists the values of $t_0$ and $t_1$ for various sensor paths. The values of $f_0$ and $f_1$ were set to \SI{100}{kHz} and \SI{200}{kHz} respectively, given that the pulse actuation frequency is \SI{150}{kHz}.

\begin{table}[!t]
\centering
\small
\caption{Time of arrival $(t_0)$ and exit $(t_1)$ of $S0$ mode for various sensor paths.}
\label{tab:TAO_manual}
\begin{tabular}{|l|l|l|}
\hline
\textbf{Sensor Path}    & $t_0$ [\SI{}{\micro s}] & $t_1$ [\SI{}{\micro s}]\\ \hline \hline
Vertical (P12, P56) & 28  & 61.3               \\ 
Horizontal (P25) & 55.9  & 89.3               \\ 
Diagonal (P15, P24, P26, P53)   & 62.5  & 95.9               \\ \hline
\end{tabular}
\end{table}

Using equations \eqref{EQ:DI1} and \eqref{EQ:DI2}, we obtained feature vector of size $2\times110$ for each of the test case datasets. The same test case datasets reported above were used for this analysis. Using these features, we created two GMMs for each path (baseline and monitoring, for each damage) and computed their KL divergence. The trends in KL divergence values for $10$ test cases for $SNR =$ \SI{4.5}{dB} are illustrated in the top row in Figure \ref{fig:SDC_2x3}. The distinction between damaged and undamaged paths is not as pronounced for the CNN+GMM method (Figure \ref{fig:SDC_3x3}). Figure \ref{fig:DI_2x3}(a) shows that for rivet hole and added mass, the damage location was correctly identified in the DI map, whereas large error was observed for notch. Thus, using GMMs in conjunction with manually computed physics-defined features ($DI_1$ and $DI_2$) is not as effective as the CNN+GMM method for countering EOC variations.

\subsection{Comparison of CNN+GMM method vs. conventional SDC algorithm using denoised signals}
\label{sec:DI_SDC}
Conventional SDC computation algorithms work extremely well in well-controlled, laboratory environments where the noise levels are very low and uniform temperature is maintained. Since the CNN used in the CNN+GMM method involves filtering the input time-series signal, in this section we analyze the performance of conventional SDC computation algorithm on denoised input signals. The input signals with added noise and temperature variation (equation \eqref{EQ:Noise}) are denoised (i.e. filtered) using moving average kernel of size $3$. The filtering operation is similar to the one performed by the CNN. 
Figure \ref{fig:denoised} illustrates a representative denoised time-domain signal for path P15. The denoised signals are then used for SDC computation using the conventional correlation coefficient deviation (CCD) definition \cite{sikdar2016identification}.
Let $S_d$ denote the signal received at receiver $(j)$, and $S_b$ denote the baseline signal. Let $t_a$ denote the time of arrival for the desired GW mode in the signal and let $t_{BW}$ denote the width of the time window corresponding to the wave mode (i.e. time bandwidth) of the desired GW mode.
The SDC coefficient for sensor path $i-j$ is defined as  measure of deviation of received signal $S_d$ from baseline signal $S_b$ and expressed as CCD:

\begin{equation}
	SDC_{CCD}{(i,j)} = 1 - CC(i,j)
\label{EQ:CCD}
\end{equation}
\begin{multline}
    CC{(i,j)} = \\
    \frac{1}{\sigma_b \sigma_d}\frac{\sum_{t_a}^{t_a+t_{BW}} (S_b - \mu_b ) (S_d - \mu_d )}{\sum_{t_a}^{t_a+t_{BW}} \sqrt{(S_b - \mu_b )^2} \sum_{t_a}^{t_a+t_{BW}} \sqrt{(S_d - \mu_d )^2}}
\end{multline}

Here, $\mu_b, \mu_d$ are mean values and $\sigma_b,\sigma_d$ are standard deviation values of the samples in baseline signal $S_b$ and received signal $S_d$, respectively.
The trends in SDC using equation \eqref{EQ:CCD} computed for the same $10$ test cases for SNR = \SI{4.5}{dB} are shown in the bottom row of Figure \ref{fig:SDC_2x3}, and corresponding DI maps computed using the mean values of $SDC_{CCD}$ are shown in Figure \ref{fig:DI_2x3}(b). This conventional method produces large errors and is thus entirely ineffective in countering EOC variations, thus validating the impact of the novel CNN+GMM method proposed in this work.
It is important to note that the range of amplitudes of SDCs may vary depending on their definition (CCD, KL-divergence etc.) However, the relative difference in the values of SDCs for damaged paths and undamaged paths determines the accuracy of damage localization. Higher contrast between values of SDCs for damaged paths and undamaged paths yields better localization results. The DI maps are normalized to the pixel with the highest DI value i.e. the highest intensity in DI maps shown in Figures \ref{fig:DI_3x3} and \ref{fig:DI_2x3} is $1$.

\section{Experimental validation}

\subsection{Experimental setup}

\begin{figure}[!t]
\centering
\includegraphics[width=0.8\linewidth]{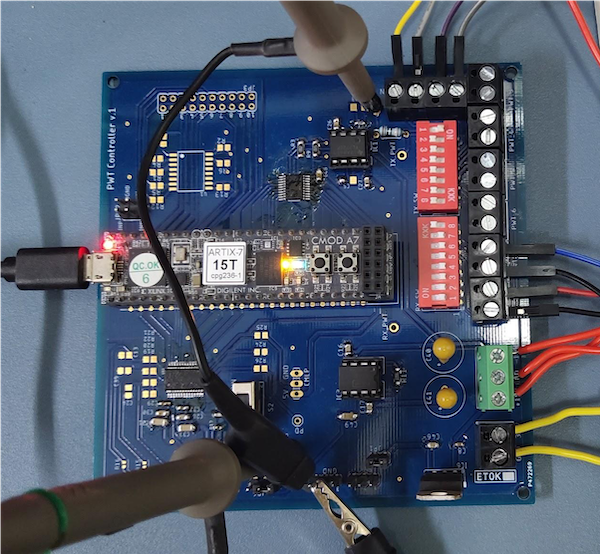}
\caption{Photograph of the PCB embedded platform for transduction and data acquisition used for performing experiments.}
\label{fig:pcb}
\end{figure}

A portable embedded system was developed for transduction of guided wave ultrasonic signals and data acquisition and storage. The system is based on a low-cost field programmable gate array (FPGA) device (Xilinx Artix\textregistered-7 XC7A15T-1CPG236C, Digilent Cmod A7-15T module), and has provision for connecting $8$ PZT channels. The FPGA is interfaced with a \SI{12}{} bit, parallel input, multiplying digital to analog converter (DAC) - Texas Instruments DAC7821, to generate the 5-cycle Hanning pulse actuation voltage centered at \SI{150}{kHz}. Since the DAC7821 is an unbuffered current output DAC, a transimpedance amplifier constructed with Texas Instruments TL072 operational amplifier IC is connected to the output of the DAC, resulting in peak-to-peak amplitude of \SI{4.8}{V} for the 5-cycle Hanning pulse. The receiver PZT output is amplified using Texas Instruments INA128 instrumentation amplifier and digitized using Maxim Integrated MAX-1426 analog to digital converter (ADC) with \SI{10}{} bit resolution and \SI{10}{Msps} sampling rate. The system is implemented on a printed circuit board (PCB), with the $8$ channels connected to the ADC and DAC using $8-$position switch banks. The system is capable of storing \SI{4096}{} samples obtained at \SI{10}{Msps} i.e. \SI{409.6}{\micro s} for each channel in the memory (Block RAM) in the FPGA. A photograph of the PCB is shown in Figure \ref{fig:pcb}.

\begin{figure}[!t]
\centering
\includegraphics[width=0.8\linewidth]{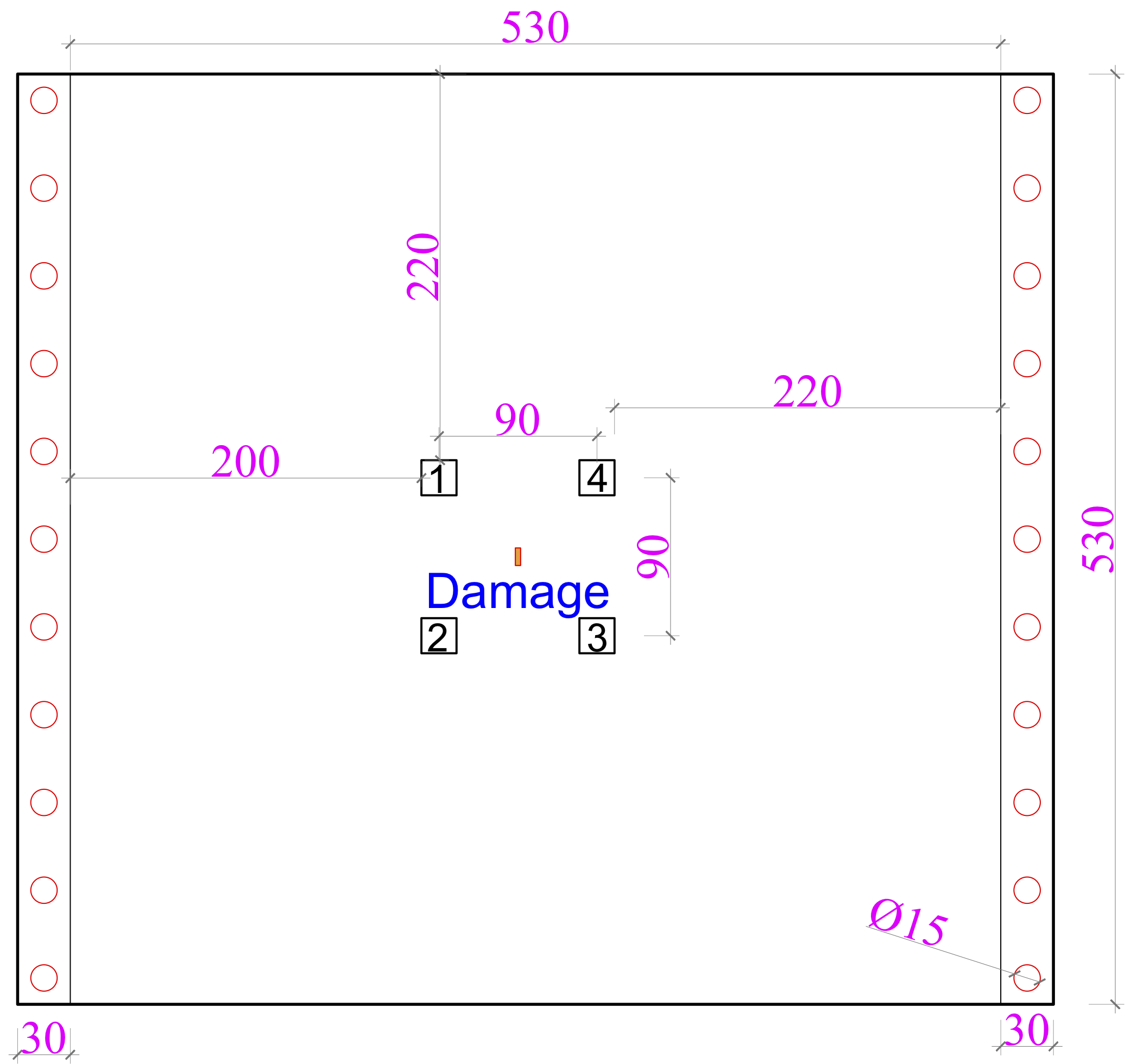}
\caption{Layout of \SI{2}{mm} thick aluminum panel with 4 PZT transducers used for experimental testing. A notch defect of dimensions \SI{10 x 3}{mm} was introduced in the panel.}
\label{fig:drawing_expt}
\end{figure}

\begin{figure}[!t]
\centering
\includegraphics[width=0.9\linewidth]{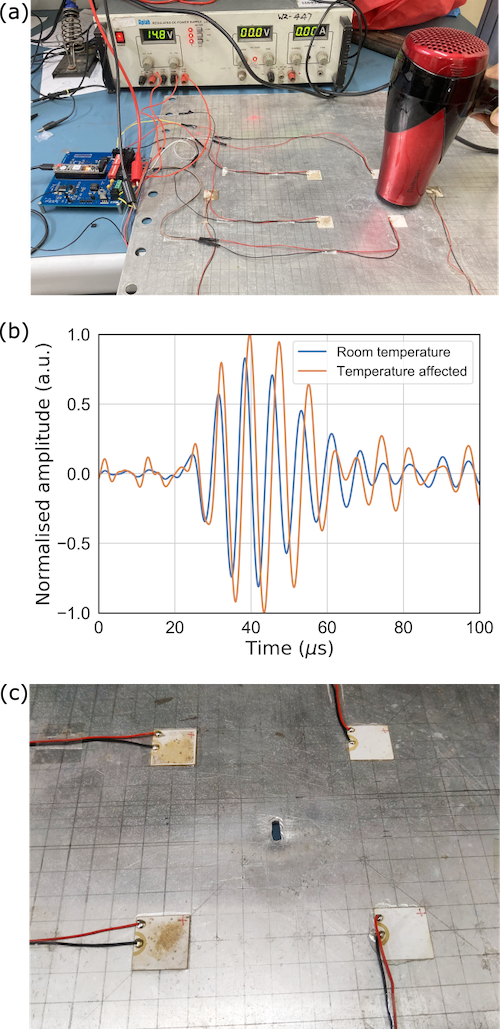}
\caption{(a) Photograph of aluminum panel (without damage) instrumented with PZT transducers connected to PCB. A hair dryer was used to heat a portion of the panel (example, above path P34 as shown) to induce non-uniform temperature profile on the panel. (b) Indicative temperature affected data for panel with notch defect at room temperature and in presence of non-uniform temperature variation for path P34. (c) Photograph of notch defect introduced in the panel.}
\label{fig:panel}
\end{figure}

\begin{figure}[!t]
\centering
\includegraphics[width=0.8\linewidth]{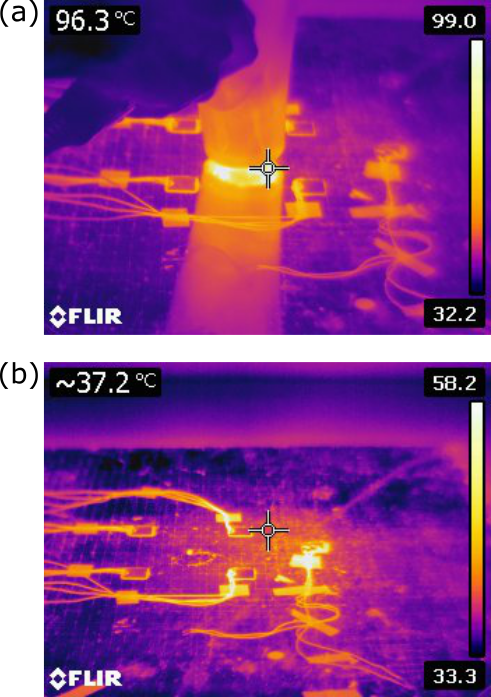}
\caption{Photographs of panel captured with thermal imaging camera. (a) Highest temperature reached during application of heat with hair dryer is approximately \SI{100}{\degreeCelsius}. (b) Non-uniform heating of the panel is noticed in the image captured few seconds after removal of heat source. The heat source was present atop the location identified by the marker in the middle of the image.}
\label{fig:thermal_FLIR}
\end{figure}

An aluminum panel of dimensions \SI{530 x 530 x 2}{mm} was used as test structure in the experiment. Four PZT-5H transducers of dimensions \SI{20 x 20 x 0.4}{mm} were attached on the panel with center-to-center separation of \SI{90}{mm} in horizontal and vertical directions, as shown in the layout in Figure \ref{fig:drawing_expt}. Baseline data for all sensor paths were recorded at room temperature on the panel without any damage. Next, $8$ data sets were collected by applying heat to a portion of the panel using a heat source (hair dryer), as shown in Figure \ref{fig:panel}(a). For each data set, the heat source was held atop an arbitrarily chosen location on the panel for up to $2$ minutes, after which the heat source was removed and the data for all paths were recorded sequentially (Figure \ref{fig:panel}(b) shows indicative data for path P34 corresponding to heat source position in Figure \ref{fig:panel}(a)). Next, a notch defect of dimensions \SI{10 x 3}{mm} was introduced in the panel using a handheld electric drill (position shown in Figure \ref{fig:drawing_expt}, photograph of defect shown in Figure \ref{fig:panel}(c)). After introduction of the notch defect, $9$ data sets were collected - one at room temperature, and $8$ data sets for various non-uniform temperature profiles using the heat source. Figure \ref{fig:thermal_FLIR} shows indicative photographs of the panel obtained with FLIR E60 thermal imaging camera during and after application of heat.

\subsection{Experimental results and discussion}

\begin{figure}[!t]
\centering
\includegraphics[width=0.8\linewidth]{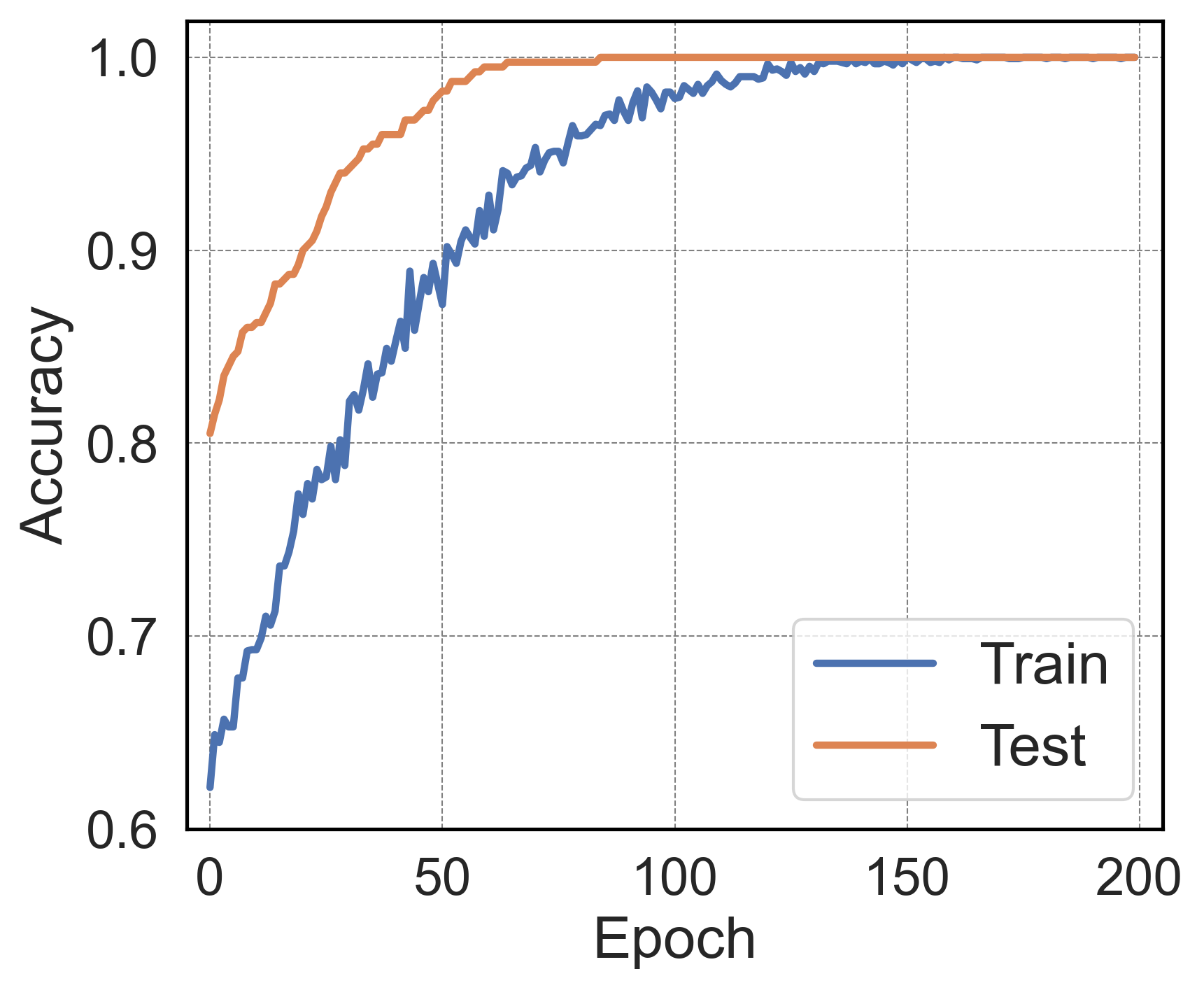}
\caption{Accuracy curves for training and testing of 1D CNN corresponding to sensor path P24 in noise-augmented experimental data set for $200$ epochs.}
\label{fig:LossCurveExp}
\end{figure}

Data for four sensor paths (P12, P13, P14, P24, for sensor positions as marked in Figure \ref{fig:drawing_expt}) was considered for analysis.
The data thus recorded were denoised using $5^{th}$ and $6^{th}$ coefficients of $7-$level wavelet decomposition, using Daubechies 10 (db10) as mother wavelet. This is essential to remove electronic offset from the circuit components on the PCB. 
Next, each of the $9$ data sets collected for undamaged panel and panel with notch defect was used to generate noise-augmented data with \SI{4.5}{dB} SNR (in a manner similar to equation \eqref{EQ:Noise}).
For each class (undamaged and notch), a total of $900$ noise-augmented time series were generated per path for training CNN for binary classification, and additional $90$ time series per path were generated for testing (monitoring).
Table \ref{tab:table_expt} shows an overview of the CNN model used for analyzing the experimental data. The model was trained for $200$ epochs, and produced perfect (\SI{100}{\%}) accuracy with test data sets. Since the sensors are placed closer to each other (\SI{90}{mm} separation along horizontal and vertical paths) on the panel used in the experiment, as compared to the FE generated data (\SI{150}{mm} and \SI{300}{mm} separation between adjacent sensors along vertical and horizontal paths respectively), we use the first \SI{100}{\micro s} portion of each time series, that contain the $S0$ mode. Accuracy curves (for $200$ epochs for path P24) are shown in Figure \ref{fig:LossCurveExp}. 

\begin{table}[!tbp]
\caption{Summary of CNN architecture for binary classification of experimental data}
\label{tab:table_expt}
\small
\begin{tabular}{p{2.75cm} | p{1.75cm} | p{2.5cm}}
\hline

\textbf{Layer (type)} & \textbf{Output shape} & \textbf{Number of parameters} \\ \hline

Conv1D & $997 \times 16$  & $64$ \\
MaxPooling1D & $498 \times 16$ & $0$ \\
Dropout & $498 \times 16$ & $0$ \\
Flatten & $7968$ & $0$  \\
Dense (ReLU) & $16$ & $127504$ \\
Dense (Sigmoid) & $2$ & $34$ \\ \hline
\multicolumn{2}{c|}{\textbf{Total trainable parameters}} & $127602$ \\ \hline
                                 
\end{tabular}
\end{table}

\begin{figure}[!t]
\centering
\includegraphics[width=0.8\linewidth]{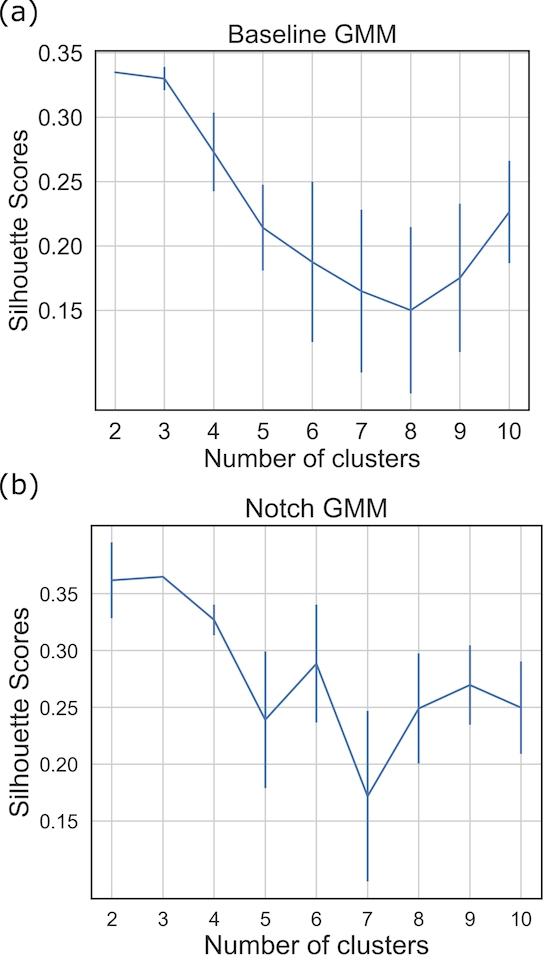}
\caption{Silhouette scores for various number of clusters for path P24 in noise-augmented experimental data set, shown for (a) baseline (undamaged panel), and (b) notch defect.}
\label{fig:SIL_exp}
\end{figure}

\begin{figure*}[!t]
\centering
\includegraphics[width=0.9\linewidth]{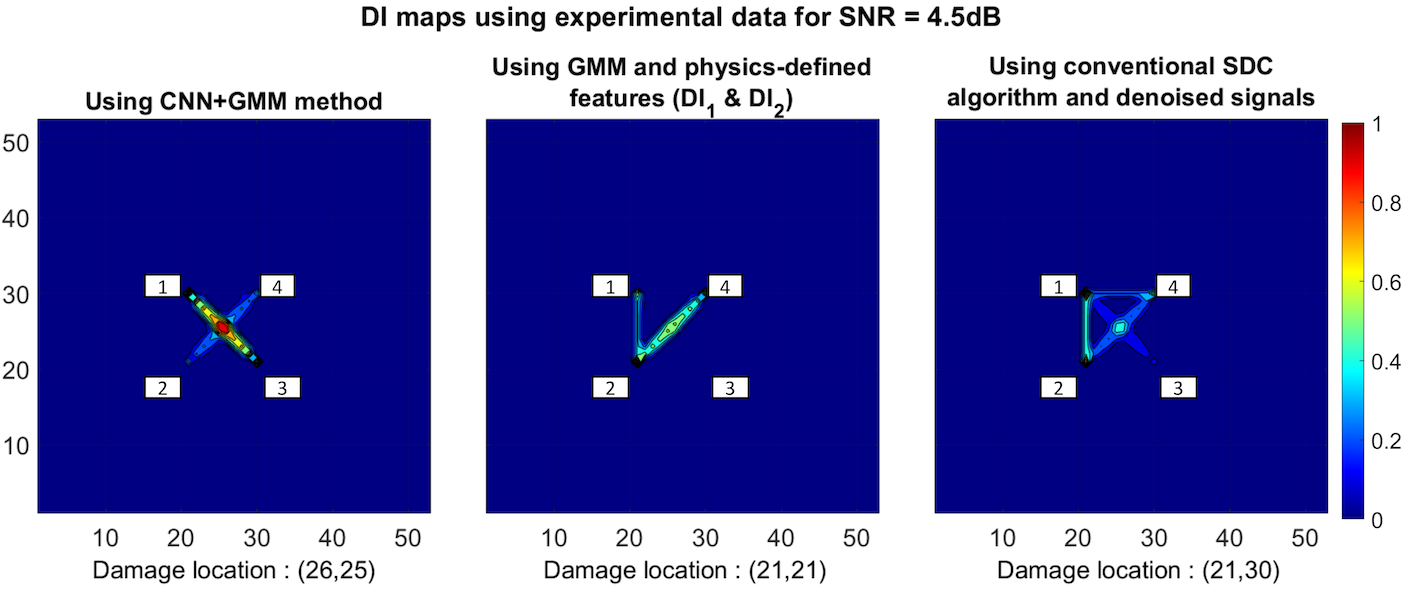}
\caption{
DI maps for noise-augmented experimental data under varying non-uniform temperature profiles on aluminum panel with notch defect, obtained using SDC based damage localization algorithm. SDCs are computed using computed using: (a) the proposed CNN+GMM method as described in equation\ref{eq:KL}, (b) KL divergence of GMMs modeled using physics-defined features $DI_1$ and $DI_2$ (equations \eqref{EQ:DI1} and \eqref{EQ:DI2}), and (c) conventional SDC definition using CCD metric (equation \eqref{EQ:CCD}) applied to signals denoised using moving average kernel of size $3$. Only the proposed CNN+GMM method is able to accurately localize the damage on the panel.
}
\label{fig:DI_exp}
\end{figure*}

Out of the $16$ features extracted from the penultimate layer of the CNN, the $2$ most prominent features (determined using PCA) were used to construct GMMs. The optimum number of clusters was determined to be $3$, based on the silhouette score (Figure \ref{fig:SIL_exp}).
KL divergence between the two GMMs was calculated as described in equation \eqref{eq:KL}. To illustrate the performance of CNN+GMM method for experimental data, comparative analysis was conducted using GMM with physics-defined features $DI_1$ and $DI_2$ as well as conventional SDC algorithm (CCD) with denoised signals as explained in subsections \ref{sec:DI1&DI2} and \ref{sec:DI_SDC},
respectively. For $90$ test samples, values of $DI_1$ and $DI_2$ were computed for baseline and notch data sets. The time of arrival and exit time for various sensor paths used for $DI_1$ and $DI_2$ computation are listed in Table \ref{tab:TAO_manual_exp}.
KL divergence between the two GMMs formed using the feature vector of size $90\times2$ for each sensor path was used for DI mapping. In case of conventional SDC algorithm using CCD, the average of CCD values of the $90$ denoised test samples for each path was considered for the computation.
SDC values for noise augmented test data using the three approaches are listed in Table \ref{tab:SDC_exp} and DI maps obtained using these SDC values are shown in Figure \ref{fig:DI_exp}. We observe that only CNN+GMM method yields higher SDC values for paths containing the damage (diagonal paths) as compared to other paths, and therefore gives correct damage location for temperature affected experimental data.

\begin{table}[!t]
\centering
\small
\caption{Time of arrival $(t_0)$ and exit $(t_1)$ of $S0$ mode for various sensor paths in experimental data set.}
\label{tab:TAO_manual_exp}
\begin{tabular}{|l|l|l|}
\hline
\textbf{Sensor Path}    & $t_0$ [\SI{}{\micro s}] & $t_1$ [\SI{}{\micro s}]\\ \hline \hline
Vertical (P12) & $16.8$  & $50.1$               \\ 
Horizontal (P14) & $16.8$  & $50.1$               \\ 
Diagonal (P13, P24)   & $28$  & $61.3$               \\ \hline
\end{tabular}
\end{table}

\begin{table*}[!t]
\centering
\small
\caption{Comparison of SDC values computed using various methods for noise-augmented experimental data. Only the CNN+GMM method shows higher SDC values for paths containing damage (diagonal paths) as compared to the vertical and horizontal paths. Numbers in parentheses indicate normalized value of SDC for each method.}
\label{tab:SDC_exp}
\begin{tabular}{|p{2.5cm}|p{4cm}|p{4cm}|p{4cm}|}
\hline
\textbf{Sensor path}  & Proposed CNN+GMM method & GMM and physics defined features $(DI_1$ and $DI_2)$ & Conventional SDC algorithm (CCD) and denoised signal\\ \hline \hline
P12 (Vertical)  & $0.02~(0.05)$  & $6.62~(0.73)$   & $0.64~(1)$        \\ 
P14 (Horizontal) & $0.05~(0.12)$  & $1.54~(0.17)$   & $0.41~(0.64)$         \\ \hline \hline
P13 (Diagonal)   & $0.41~(1)$  & $1.44~(0.16)$   & $0.24~(0.38)$           \\ 
P24 (Diagonal)   & $0.17~(0.41)$  & $9.11~(1)$   & $0.39~(0.61)$         \\ \hline
\end{tabular}
\end{table*}

\section{Conclusion and future work}

The proof of concept implementation of the proposed method is illustrated for GW-SHM system consisting six-sensor network mounted atop an aluminum plate, with FE simulation based data generated at various temperatures from \SI{0}{\degreeCelsius} to \SI{100}{\degreeCelsius}, with added white noise and pink noise with SNR varied from \SI{2}{dB} to \SI{10}{dB}, and experimental data with non-uniform temperature variation and added noise. We have conclusively demonstrated that CNN-based automated feature computation and analysis using KL divergence of GMMs of features to represent SDC gives far superior performance as compared to conventional methods reported previously in literature, using both simulated as well as experimental data sets. Furthermore, the CNN-based approach requires no domain knowledge (wave velocity, impact of damage on amplitude and phase of wave modes etc.) and is therefore applicable broadly for various structures.
To illustrate the efficacy of the method for damage classification and localization, the proposed algorithm was tested for different types of defects (rivet hole, added mass and notch), albeit with the same defect location, at the intersection of two diagonal sensor paths in the network. The experimental validation with notch defect also required the damage location to be at the intersection of the diagonal paths. This constraint was necessary, since the SDC based damage localization algorithm used to compute DI maps has limited spatial resolution and requires a grid of sensors. This method is most sensitive to damages that are present at the intersection of ellipses corresponding to different sensor paths. In future work, we will explore alternate methods to improve the spatial resolution of the damage localization using methods such as outer-tangent method \cite{8922690} and delay and sum algorithm \cite{MICHAELS2007482}, and also by employing deep learning coupled with regression. Availability of extensive labeled data significantly limits the applicability of supervised learning algorithms. Therefore we also seek to explore unsupervised learning and transfer learning based approaches that have more applicability and warrant further research. Additionally, we seek to also explore development of light-weight deep learning models inspired by TinyML framework \cite{sanchez2020tinyml}, that could be directly embedded in the FPGA based embedded system presented in this work, to truly design portable smart SHM systems.

\section*{Acknowledgment}

This work was partially supported by Indian Space Research Organization (ISRO) [grant no. RD/0118-ISROC00-006]. The authors thank Ms. Sahar Almahfouz Nasser and Prof. Amit Sethi at IIT Bombay for insightful discussions on deep learning, and Mr. Rambabu Sugguna at National Centre for Photovoltaic Research and Education (NCPRE), IIT Bombay for assistance with thermal imaging.

\section*{Declaration of interests}
The authors declare that they have no known competing financial interests or personal relationships that could have appeared to influence the work reported in this paper.


\bibliography{main}
\end{document}